\documentclass[11pt,american,english]{article}
\usepackage[latin9]{inputenc}
\usepackage{array}
\usepackage{float}
\usepackage{algorithm2e}
\usepackage{amsmath}
\usepackage{amssymb}
\usepackage{graphicx}

\makeatletter

\usepackage{latexsym}
\usepackage{graphicx}

\makeatother

\usepackage{babel}
\begin{document}
\title{On the validation of complex systems operating in open contexts}
\author{Alexander Poddey\thanks{Bosch Center for Artificial Intelligence, Robert Bosch GmbH, Corporate Research Campus, 71272 Renningen, Germany,  alexander.poddey@de.bosch.com} \and Tino Brade\thanks{Vehicle Safety and Assistance Systems, Robert Bosch GmbH, Corporate Research Campus, 71272 Renningen, Germany} \and Jan E. Stellet\footnotemark[2] \and Wolfgang Branz\footnotemark[2] }
\date{\the\year}

\maketitle
\begin{abstract}In the recent years, there has been a rush towards
highly autonomous systems operating in public environments, such as
automated driving of road vehicles, passenger shuttle systems and
mobile robots. These systems, operating in unstructured, public real-world
environments (the operational design domain can be characterized as
open context) per se bear a serious safety risk. The serious safety
risk, the complexity of the necessary technical systems, the openness
of the operational design domain and the regulatory situation pose
a fundamental challenge to the automotive industry. Many different
approaches to the validation of autonomous driving functions have
been proposed over the course of the last years. However, although
partly announced as \textquoteleft the solution\textquoteright{} to
the validation challenge, many of the praised approaches leave open
crucial parts. To illustrate the contributions as well as the limitations
of the individual approaches and providing strategies for 'viable'
validation and approval of such systems, the first part of the paper
gives an analysis of the fundamental challenges related to the valid
design and operation of complex autonomous systems operating in open
contexts. In the second part, we formalize the problem statement and
provide algorithms for an iterative development and validation. In
the last part we give a high level overview of a practical, holistic
development process which we refer to as systematic, system view based
approach to validation (in short sys$^{2}$val) and comment on the
contributions from ISO26262 and current state of ISO/PAS 21448 (SOTIF).\end{abstract}

\section{Introduction}

In the recent years, there has been a rush towards highly autonomous
systems operating in public environments, such as automated driving
of road vehicles, passenger shuttle systems and mobile robots. These
systems, operating in unstructured, public real-world environments
(the \emph{operational design domain} can be characterized as \emph{open
context}) per se bear a serious safety risk. The serious safety risk,
the complexity of the necessary technical systems, the openness of
the operational design domain and the regulatory situation pose a
large challenge to the automotive industry. 

Due diligence is necessary in development, release and even post release
operation, which are all related to validation aspects. Successful,
ongoing demonstration of safe operation and strict avoidance of fatal
incidents in all day use is necessary for societal acceptance. Otherwise,
a \textquoteleft winter of autonomous systems\textquoteright{} (\cite{Shalev-Shwartz2017})
might come down and the large investments taken e.g. in the automotive
industry will not pay-off.

Many different approaches to the validation of autonomous driving
functions have been proposed over the course of the last years. There
have been discussions, among others, about real world driving based
statistic approaches (following the state of the art in assisted driving),
simulation, formal patterns, partly as silver bullet (one for all
solutions), partly in combination. However, although partly announced
as \textquoteleft the solution\textquoteright{} to the validation
challenge, many of the praised approaches leave open crucial parts.
In order to be able to illustrate the contributions as well as the
limitations of the individual approaches, we give an analysis of the
fundamental challenges related to the valid design and operation of
complex autonomous systems operating in an open context in the following.
Strategies for 'viable' validation and approval of such systems can
then be discussed, based on this.

In section \ref{sec:Problem-description}, we introduce basic terms
and concepts as well as a detailed presentation of the validation-
and approval related necessities for complex systems operating in
an open context, summed up as \emph{validation challenge}. \textbf{\emph{Boldface}}
characters are used for term definitions, whereas \emph{italic} characters
indicate the use of already defined terms within the document. Section
\ref{sec:Problem-formalization} formalizes the problem statement
and provides algorithms for iterative development and validation.
In section \ref{sec:contribAll} we give a high level overview of
a practical, holistic development process, which we refer to as systematic,
system view based approach to validation (in short sys$^{2}$val)
and comment on the contributions from ISO26262 and current state of
ISO/PAS 21448 (SOTIF) for automotive focused applications.

\section{Problem statement \label{sec:Problem-description}}

\subsection{Conception of validity of complex systems in open contexts\label{subsec:Conception-of-validity}}

Autonomous systems operating in public environments are usually designed
to take over typical human tasks like e.g. driving road vehicles.
The technical systems, or strictly speaking its developers and distributors,
thereby do not only need to fulfill the tasks from a functional point
of view. They also need to take over responsibility for safe operation
and mitigation of hazardous situations, traditionally incurred by
the human operator. This is significantly different from e.g. assisted
driving where a quite limited assistance function is continuously
supervised by the driver. Over and above, the unstructured real-world
operational design domain can be characterized as an \textbf{\emph{open
context}}\index{open context}. It bears infinitely many characteristics,
possible interactions and effects, which cannot be expressed formally
complete (we refer to this as \textbf{\emph{\ensuremath{\infty}-complexity}}\index{infinte complexity}\index{infty-complexity@$\infty$-complexity}).
Moreover, the context develops in time (it is evolving) with so far
unseen characteristics and interactions appearing suddenly, e.g. post
release. Due to the complexity of the autonomous system (e.g. compared
to an assistance function, see also \emph{emergent behavior} discussed
later on) and the resulting numerous possible interactions, changes
in context might lead to undesirable behavior of the system. Both,
the complexity of the system and the context therefore require several
topics to be addressed in a systematic and holistic approach in order
to achieve a valid product. With \textbf{\emph{valid system}}\index{valid system},
we refer to a product which bears no \textbf{\emph{unreasonable risk}}\index{unreasonable risk}\footnote{According to IOS26262: Risk judged to be unacceptable in a certain
context according to valid societal moral concepts.} to users and the society, and, however subordinated, no unreasonable
risk to the manufacturer, which is related to liability issues and
costs e.g. due to unreasonable development expenditures. The following
list provides an overview of these topics: 
\begin{itemize}
\item functional safety (e.g. IEC61508, ISO26262)
\item safety of the intended functionality (SOTIF)
\item safety of machine learning
\item safety of use
\item security 
\item product liability 
\item meeting customers and society's expectations 
\end{itemize}
As will be discussed in the following, complex systems operating in
an open context will never be \textbf{\emph{perfectly valid\index{perfectly valid}}}
(i.e. a system which bears absolutely no risk). 

In accordance with system-view based approaches (e.g. STAMP \cite{Leveson2011}),
all the aspects illustrated in the above listing are part of the \textbf{\emph{high
level goal}}\index{high level goal}. The negative effect of non-achievement
of aspects of the high level goal is referred to as \textbf{\emph{loss}}\index{loss}.
Losses large enough to undermine the achievement of the overall goal
are referred to as \textbf{\emph{unacceptable loss\index{unacceptable loss}.}}
A system possibly leading to \emph{unacceptable losses} is invalid,
as it bears unreasonable risk. 

The differentiation between \emph{loss} and \emph{unacceptable loss}
is related to the (societal) \emph{tolerated risk}. For the moment,
it is sufficient to note that the (societal) tolerated risk is a complex
topic and even a moving target, which, among others, depends on attributes
like type of cause of \emph{loss} (e.g. force major cause, random
effect - like e.g. random hardware errors - or systematic malfunction)\emph{.}
Societal factors play an important role for what might be regarded
as tolerated. Societal expectations, influenced by advertising, current
jurisprudence and trends however are hard to quantify, which makes
arguing about \emph{tolerated risk} a fuzzy topic. In addition, the
validation of such systems can no longer be solved in a \textquoteleft once
for all time\textquoteright{} approach, as it will be discussed later
on in the context of \emph{ongoing validation}. 

At first we will introduce some additional terms and concepts related
to the root challenge of validation. 

\subsection{Basic concepts from development approaches\label{subsec:Basic-concepts-devApproaches}}

Over the last decades, the challenges in developing complex systems
have been approached in many different ways (we comment on the main
approaches in \ref{sec:Contribution-of-E}). In the following, we
introduce some basic concepts and terms necessary for the problem
statement. 

\begin{figure}[H]
\begin{centering}
\includegraphics[viewport=0bp 230bp 960bp 540bp,clip,width=0.9\columnwidth]{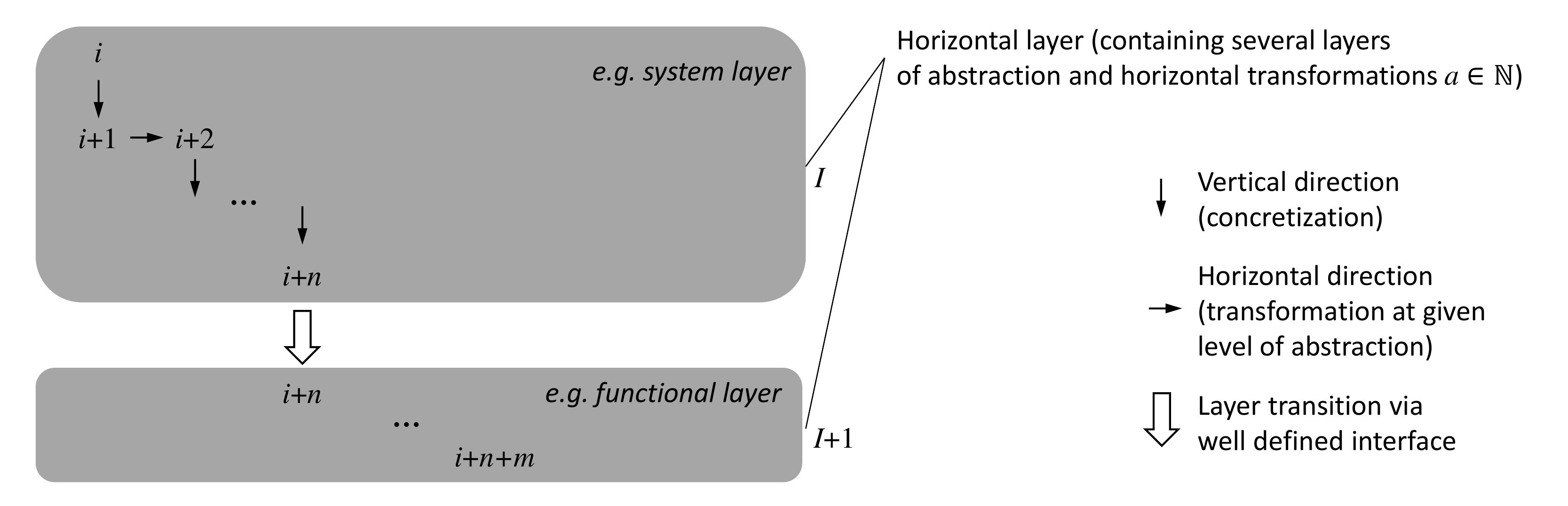}
\par\end{centering}
\caption{\label{fig:horicontalVertical}Illustration of the concepts of coarse
\emph{horizontal layers} (gray boxes) encapsulating fine-grained layers
of abstraction (e.g. $i\downarrow i+n$ in the \emph{vertical direction,}
and $i+1\rightarrow i+2$ in the\emph{ horizontal direction} in the
first box). }
 
\end{figure}

In general, the development of complex systems deals with different
layers of abstraction. Established approaches usually apply a quite
coarse definition of \textbf{\emph{horizontal layers}}\index{horizontal layer},
such as e.g. system-, functional- and implementation layer, as illustrated
in figure \ref{fig:horicontalVertical} as boxes. These horizontal
layers each encapsulate several (fine-grained) layers of abstraction
(e.g. $i\downarrow i+n$ in the first box of figure \ref{fig:horicontalVertical}).
Moving down the layers of abstraction corresponds to the \textbf{\emph{vertical
direction\index{vertical direction}}}. The basic idea behind the
usage of coarsely defined \emph{horizontal layers} is to encapsulate
the specific necessities of the different layers with well defined
interfaces and transitions among them (large arrow in the figure,
e.g. handing over a set of requirements from the system- to the functional
layer). 

Reducing complexity and supporting reuse within a given layer is approached
by encapsulation into elements (e.g. system components within the
system layer) that interact via well defined interfaces. This is referred
to as encapsulation along the \textbf{\emph{horizontal direction}}\index{horizontal direction}.
Typically, during development, there are several transformations taken
also along the horizontal direction (indicated as $i+1\rightarrow i+2$
in the first box of figure \ref{fig:horicontalVertical}). 

\subsection{Validation triangle }

\begin{figure}[H]
\begin{centering}
\includegraphics[viewport=210bp 220bp 930bp 577bp,clip,width=0.5\columnwidth]{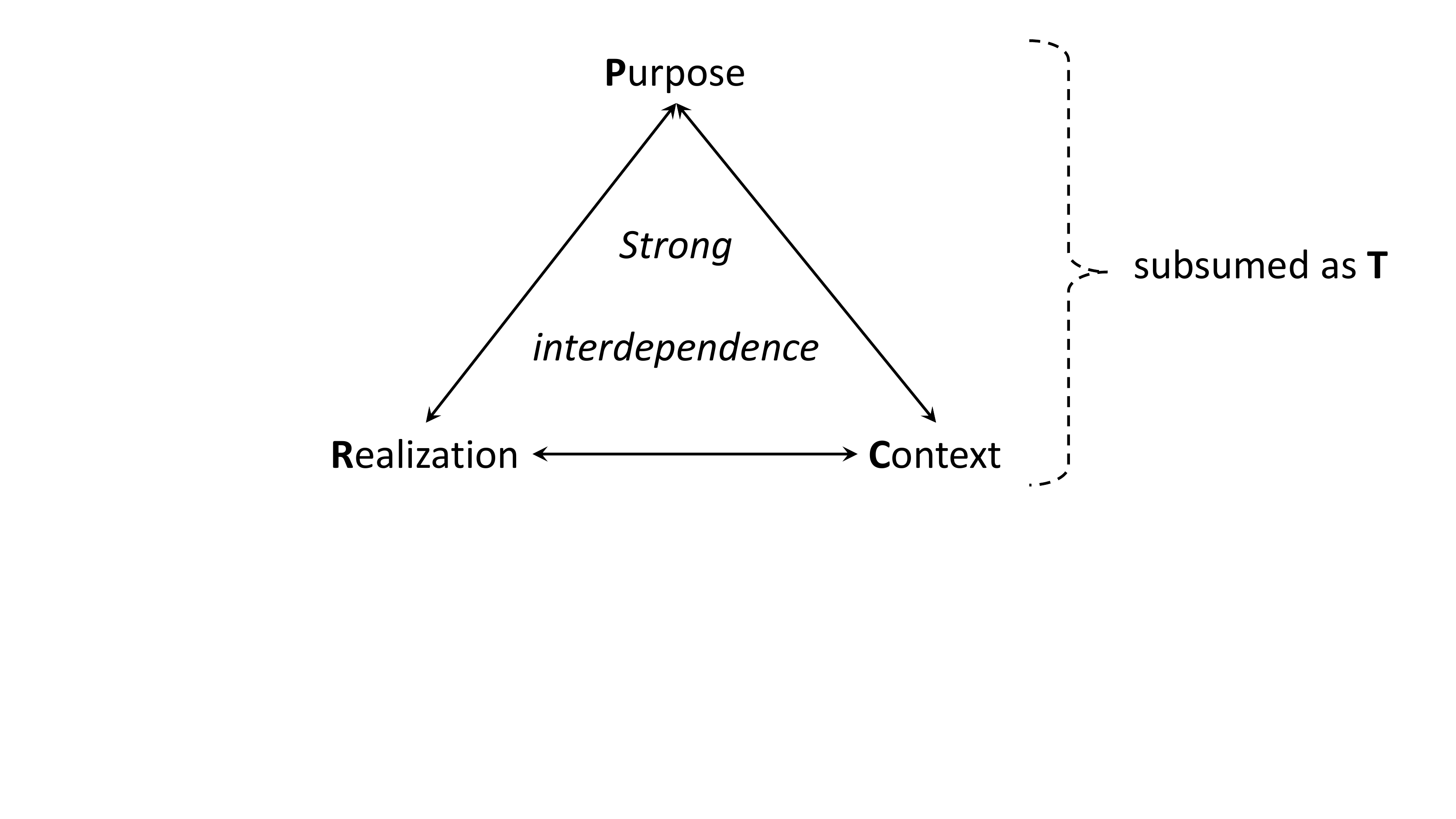}
\par\end{centering}
\caption{\label{fig:Triangle1}The \emph{validation triangle} consists of the
three, interdependent aspects purpose (P), context (C) and realization
(R), subsumed as the triade T.}
\end{figure}

Figure \ref{fig:Triangle1} provides an illustration of the three
basic aspects related to the overall validation problem (we refer
to this as \textbf{\emph{validation triangle\index{validation triangle}}}).

In general, providing a valid solution to a given task is related
to (depending on the \emph{horizontal layer}) developing, designing
or implementing a \textbf{\emph{realization R\index{R}\index{realization}}}
(e.g. a technical system) for a \textbf{\emph{purpose P\index{P}}}\index{purpose}\textbf{\emph{
}}(e.g. autonomous driving on a highway), which needs to be fulfilled
in a specific \textbf{\emph{context C\index{C}}}\index{context}
(e.g. on german highways at daytime). This three-way dependency is
present across all levels of abstraction, e.g. on the highest level
as in the above example, down to the implementation of a concrete
sub-function or technical component. The technical component as well
is built to fulfill an intended purpose (contributing to the main
purpose of the technical system) in a given context. We use the shorthand
notation \textbf{\emph{\index{T}T}} to refer to the interdependent
\textbf{\emph{triad\index{triad}}} of $P,\,R,\,C$. 

In the following, we will step-wise expand figure \ref{fig:Triangle1}
to build up an illustration of the manifold and interrelated aspects
of the actual \emph{validation problem} related to complex systems
operating in open contexts.

\subsection{Implicit infinite complexity \label{subsec:Implicit-infinite-complexity}}

The fundamental challenge of validation of complex systems operating
in an open context is that none of the three aspects of the\emph{
validation triangle} - \emph{purpose (P)}, \emph{context (C)}, and
\emph{realization (R)} - can be expressed in a formally complete manner
and, as already discussed above, the expression of one aspect even
depends on the remaining two. This is illustrated in figure \ref{fig:Triangle3new}a)
as a fuzzy black frame. In the following, the cause of this will be
discussed per aspect.

On the one hand, as already discussed above, unstructured real-world
operational design domains (open contexts) bear infinitely many possible
interactions and effects, they are \emph{\ensuremath{\infty}-complex}.
The \emph{purpose}, on the other hand, is based on implicit expectations,
which also cannot be expressed formally complete. Over and above,
statements regarding a certain \emph{purpose} depend on the related
\emph{context} of application and possibly even on the characteristics
of the chosen \emph{realization} and therefore most often are based
on implicit \emph{assumptions} about these aspects of the \emph{validation
triangle}. 

Regarding the purpose, we differentiate between the \textbf{\emph{aimed
purpose\index{aimed purpose}}} (implicitly expected), which is necessarily
vague, and the\textbf{\emph{ intended purpose\index{intended purpose},}}
relating to explicitly expressed expectations (e.g. a specification).
We apply here the meaning of \textquoteleft intended\textquoteright{}
as \textquoteleft explicitly expressed\textquoteright{} as applied
in the context of ISO26262 \& SOTIF in connection with\textbf{\emph{
intended functionality (I)}}\footnote{\emph{Terms defined by ISO26262 / SOTIF are annotated by (I) in ordert
to explicitly indicate this circumstance.}\textbf{\emph{ }}} (behavior specified for an item, system, or element excluding safety
mechanisms) and \textbf{\emph{intended behavior (I)}}\footnote{Don\textquoteright t be irritated by the apparent circular dependency
between the two definitions via the term behavior. Specified behavior
in the definition of the intended functionality is meant as \quotesinglbase what
was specified regarding the functionality\textquoteleft . }\textbf{\emph{\index{intended behavior}}} (specified behavior of
the intended functionality including interaction with items). The
difference between \emph{intended} (explicitly expressed) and \emph{aimed}
(implicitly expected) can be illustrated as the difference between
what has been stated and what was actually meant. One of the challenges
related to complex systems in open contexts is that the developer,
the customer or even the society might be able to express an expectation
e.g. about appropriate behavior of the system for a given, specific
\textbf{\emph{setting\index{setting}}} (related to a specific\emph{
realization} in a specific \emph{context}). However, due to the \emph{\ensuremath{\infty}\textendash complexity},
it is non-trivial or rather impossible to identify all relevant \emph{settings}.
The explicit expectations therefore are usually expressed quite abstract
leaving open room for (\emph{setting} specific) interpretation. The
same explicit but abstract statement might even lead to quite different
interpretations, depending on different concretizations of the \emph{setting.} 

Vice versa, the necessary mapping of the \emph{\ensuremath{\infty}\textendash complex}
\emph{context} onto a reduced subset of \textquoteleft expected to
be relevant\textquoteright{} \emph{context} (valid projection onto
a finite \emph{representation}) strongly depends on the related specific\emph{
purpose} and\emph{ realization}. For example, the presence and characteristics
of metallic structures might be relevant for radar-based\emph{ realizations},
whereas thermic radiation might be irrelevant, as long as no corresponding
sensing system is applied, or possible irritation of the \emph{realization}
due to thermic radiation might happen. This illustrates that the \textbf{\emph{representativeness\index{representativeness}}}
(i.e. a subset of something accurately reflects the larger super-set),
which is a basic precondition for statistic argumentation, cannot
easily be achieved and is impossible to achieve for a complex system
in evolving open contexts without a valid analysis of all aspects
of the \emph{validation triangle} (this is referred to as \textbf{\emph{representativeness
challenge}}). 

Last but not least, the same field of tension is present between the
expected and true/effective characteristics and behavior of a \emph{realization}.
The properties of complex systems - independent of the effect of the
operation in an open context - are \textbf{\emph{emergent}}\index{emergent}.
I.e. the effective properties and behavior of a realization is a result
of the complex interplay of the components and can therefore, again,
not easily be expressed formally complete. We refer to this as \textbf{\emph{\index{emergent characteristics}emergent
characteristics}} and \textbf{\emph{emergent behavior}}\index{emergent behavior}.
This is a second aspect, on top of the complex interaction between
the realization and the context. As a consequence, an explicitly expressed
\emph{expected behavior} of a \emph{realization} (related to the \emph{intended
behavior} (I) of SOTIF) might be different from the \emph{effective
behavior} as well as different from the \textbf{\emph{\index{effectively necessary behavior}effectively
necessary behavior}} \textendash{} necessary to meet the \emph{purpose}.
We will introduce a special notation for the expected and effective
apects of the triade in section \ref{subsec:Model-of-reality}.

\subsection{Subspace of perfectly valid solutions $\tilde{T}$ }

As discussed in the previous section, the space of possible triad
combinations T is infinitely large and can not explicitly be expressed.
However, in the sense of a thought experiment, suppose an omniscient
observer could classify any possible triad either as \emph{perfectly
valid} or not. By this, he could thereby explicitly determine the
\textbf{\emph{decision boundary}}, which reduces both, the \textbf{\emph{relevant
hyper volume}} in $T$-space to a finite volume and the implicitly
regarding the\emph{ decision boundary} to zero. The subspace of \emph{perfectly
valid} triads will be denoted by the tilde notation $\tilde{T}$. 

\begin{figure}[H]
\begin{centering}
\includegraphics[viewport=15bp 235bp 538bp 960bp,clip,width=0.9\columnwidth]{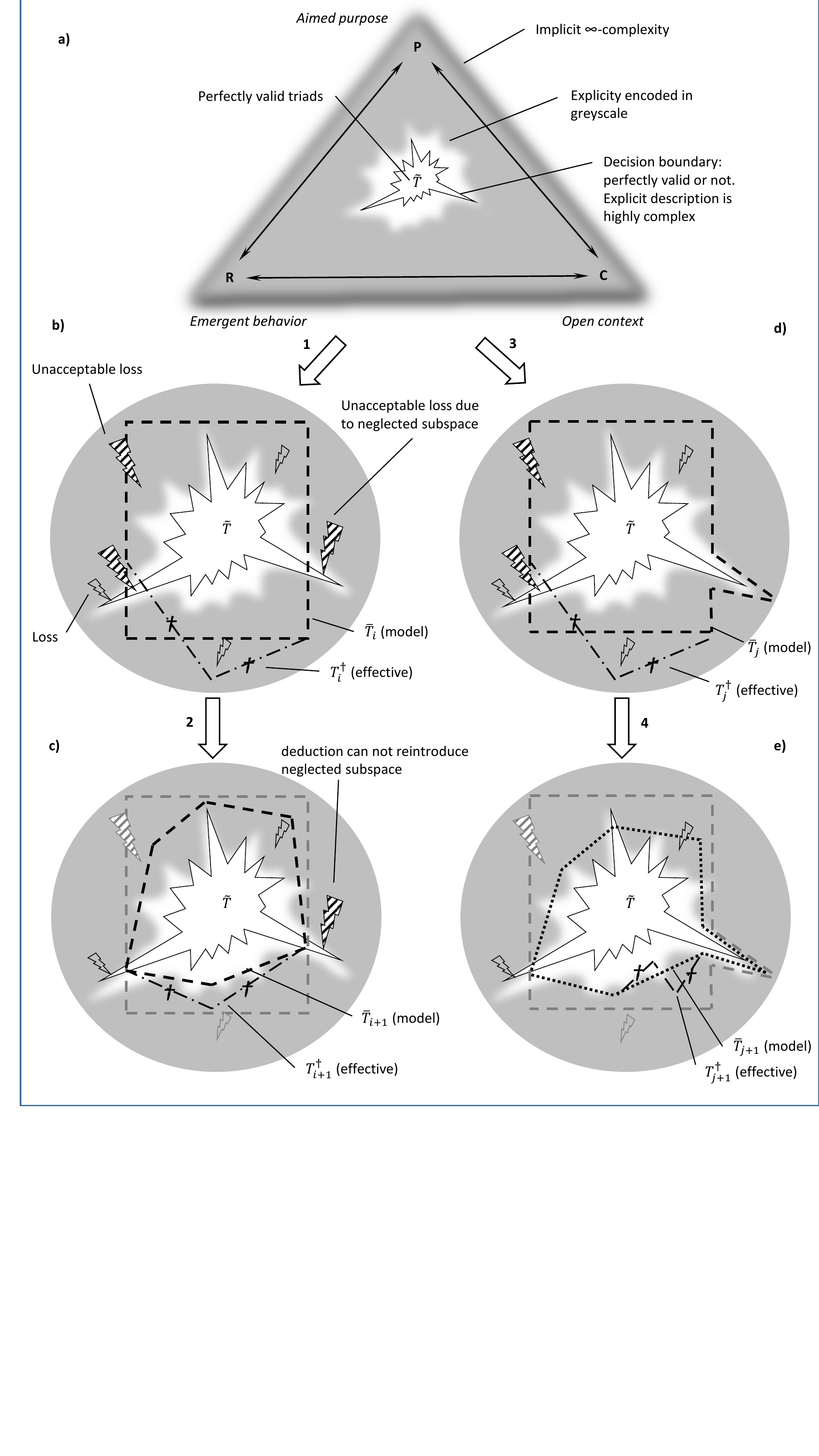}
\par\end{centering}
\caption{\label{fig:Triangle3new}The implicit $\infty$-complex space of possible
triad combinations (fuzzy black frame in a), the subspace of \emph{perfectly
valid solutions} $\tilde{T}$ and the effect of alternative deductions,
aiming at approximating $\tilde{T}$ (illustrated in b-e). For further
details see section \protect\ref{subsec:Development,-iterative-deduction}. }
\end{figure}

Having an explicit description of the decision boundary would reduce
the validation problem effectively to a verification task (i.e. checking
a given triad $T$ against a fully explicit decision boundary for
$\tilde{T}$). However, the description of the decision boundary will
necessarily be highly complex, as illustrated in figure \ref{fig:Triangle3new}
by the ragged curve. In reality, the decision boundary could neither
be determined nor would the explicit complexity be manageable. In
anticipation of section \ref{subsec:The-development-goal}, due to
this, the developement goal is targeted at a \emph{valid system} (i.e.
a sytem bearing no unreasonable loss, denoted by $\tilde{T}+loss$),
not a \emph{perfectly valid system} ($\tilde{T}$). 

\subsection{Model of reality $\bar{T}$\label{subsec:Model-of-reality}}

From section \ref{subsec:Implicit-infinite-complexity}, it becomes
clear that developing complex systems for open contexts necessarily
deals with more or less simplified \emph{representations} of the \emph{\ensuremath{\infty}-complex}
reality (including mutual interaction) for all three aspects (\emph{purpose},
\emph{context} and \emph{realization}) across all levels of abstraction.
According to the most general definition, we refer to this \emph{representation}
of reality as \textbf{\emph{model\index{model}}}. Just as argued
above, a model can never be generally complete \& correct (i.e. valid)
in every aspect. It can only be sufficiently complete \& correct for
a certain purpose in a given context. Specifically, we are interested
in models of \emph{valid systems} (i.e. models of valid triads). 

Our work is based on \cite{Spanfelner2014}, who analysed the role
of models in the context of ISO26262. We generalize this to the \emph{triangle
of validation} and hence to the more general task of validation of
complex systems operating in an open context. 

Building a model is a form of \textbf{\emph{deduction\index{deduction}}}
(sometimes also referred to as \textbf{\emph{concretizations}}\index{concretizations}),
which might happen implicitly e.g. in the form of a mental model unconsciously
applied by the developer or explicitly e.g. in the form of a formal
model, a written requirement or even written code. It is important
to note that all statements, reasoning and arguing about any aspect
of the \emph{validation triangle} is, as emphasized by the previous
sections, necessarily related to a (more or less reduced) \emph{model}
of the reality. 

We denote this \emph{model} of reality by the 'bar' notation, e.g.
$\bar{T}$ for the \emph{models} related to the aspects of the \emph{validation
triangle}, specifically $\bar{P}$, $\bar{R}$ and $\bar{C}$. Due
to the \emph{\ensuremath{\infty}\textendash complexity}, the true
characteristics, in the following denoted by the dagger notation,
e.g. $T^{\dagger}$, can not be explicitly expressed at any level
of abstraction. In addition we include the interdependence of the
aspects into the notation, so that:
\begin{itemize}
\item For a given \emph{realization R}, we are able to think, argue, reason
about the related properties and behavior only in \emph{representations}
of reality, i.e. models $\bar{R}(\bar{C},\bar{P})$, which are not
necessarily consistent with the true characteristics and behavior
$R^{\dagger}(C^{\dagger},P^{\dagger})$. 
\item Regarding the \emph{context C}, the expected to be relevant part $\bar{C}(\bar{R},\bar{P})$
might not necessarily be consistent with the effectively relevant
one $C^{\dagger}(R^{\dagger},P^{\dagger})$.
\item The mapping of the \emph{aimed purpose P} in form of $\bar{P}(\bar{C},\bar{R})$
might not necessarily meet the effectively achieved purpose $P^{\dagger}(C^{\dagger},R^{\dagger})$.
\end{itemize}
In figure \ref{fig:Triangle3new}, we indicate the consecutively derived
models of reality $\bar{T}$, corresponding to the developement process,
as dashed lines. The related effective triangle $T^{\dagger}$ as
dash-dot-dagger line. 

\subsection{Deductive gap, validation and evidence\label{subsec:Deductive-gap,-validation}}

Building a \emph{model} is a form of \emph{deduction}\footnote{Note that we apply the term deduction of model for both, deductive
and inductive model building. The effect of inductive model building
(i.e. deriving a model based on specific observations of reality),
namely a gap between the resulting model and the reality, as well
as the related necessary validation is comparable to the deductive
gap. }. Different types of \emph{deductions} appear all along the process
from the initial formulation of a product idea to the implementation
of a concrete product. An example from the top-most level, as discussed
above, would be textually writing down the \emph{intended purpose},
the relevant aspects of the \emph{context} and the characteristics
of the \emph{realization} (e.g. as set of requirements and constraints).
Other examples would e.g. be the semantic transformation of textual
\emph{representations} (e.g. requirement transformation like decomposition
and refinement), transformation of textual \emph{representations}
to more formal / testable \emph{representations}, system- or functional
decomposition, model reduction and transition from specification to
concrete implementation. 

The unavoidable \emph{deductions} however are at the heart of the
validation problem. Following from the above discussion, it is clear
that every model building is necessarily based on explicit and most
often even implicit \emph{assumptions}. In other words, the deduced
\emph{representation} corresponds to the initial \emph{representation}
only under certain \textbf{\emph{assumptions}}\index{assumptions}.
If these \emph{assumptions} are not justified (even temporarily),
the deducted \emph{representation} is invalid, and the derived \emph{realization}
provides insufficient characteristics compared to both, the\emph{
intended} and \emph{aimed purpose} of the initial \emph{representation}
for the relevant \emph{context}. In other words, there possibly exists
a gap between the initial- and the effect of the deducted \emph{representation}.
We therefore refer to this as \textbf{\emph{\index{deductive gap}deductive
gap}} and \textbf{\emph{insufficiency\index{insufficiency}}} of the
deducted \emph{representation}. 

The complex multi-dependency of the related aspects, as well as the
impossibility to formally complete express them is characteristic
for the \emph{validation problem}. In contrast to that, \textbf{\emph{verification}}
\index{verification}is related to checking for differences between,
at least in principle, formally complete expressible \emph{representations}.
The adverse effect of the \emph{deductive gap} is quite problematic.
Due to the necessary \emph{deductions}, the system might operate as
specified (the specification itself is the result of several \emph{deductions})
and the implemented system might be well verified against this specification,
but, nevertheless, it might not succeed to fulfill neither the\emph{
intended} nor \emph{aimed purpose}, showing malfunctioning behavior
leading to unsafe operation. In contrast to e.g. random hardware errors,
the effect of the \emph{deductive gap} is systematic, possibly leading
to a systematically harmful system, which is intolerable. SOTIF partly
addresses this, as discussed in section \ref{sec:ContributionsIsoSotif}. 

In addition, the for complex systems unavoidable difference between
the expected ($\bar{T}$) and the effective ($T^{\dagger}$), related
to the deductive gap, might have a strongly adverse effect on the
validation activity (i.e. hypothesis checking). This is because usually,
the same projections of reality underlying $\bar{T}$ are typically
applied during the derivation of validation tasks and analysis of
results. This leads to the fact that the results of a hypothesis check,
e.g. a test, apparently support the hypothesis, although it is, in
fact, wrong. This has a much more adverse effect than insufficient
or missing evidence (related to an\textbf{\emph{ unfounded argument}},
see also \cite{Heikkilae2017}). We therefore refer to this effect
as \textbf{\emph{misleading argument}}. 

For some of the many different types of gaps resulting from different
deduction types mentioned above, specific names are used. (see e.g.
\cite{Spanfelner2014}). We will introduce here only two frequently
used terms, namely \textbf{\emph{semantic gap}}\index{semantic gap},
which relates to the gap following from the deduction of linguistic
\emph{representations} as well as \textbf{\emph{specification gap\index{specification gap}}}
relating to the gap between the effect ($T^{\dagger}$) of a concrete
specification ($\bar{T}$) (subsuming several steps of \emph{deduction},
including the transition from the \emph{\ensuremath{\infty}-complexity}
to a finite \emph{representation}) and the implicitly aimed (i.e.
a valid soultion,$\tilde{T}$). 

Over the course of the last years, there have also been proposed different
terms for the \emph{adverse effect of the deductive gap} (i.e. the
\emph{insufficiency}), e.g. \textbf{\emph{\index{functional insufficiency}functional
insufficiency (I)}}\footnote{Term used in the context of Safety Of The Intended Functionality (SOTIF),
in particular in the definition of SOTIF as 'absence of unreasonable
risk due to hazards resulting from functional insufficiencies of the
intended functionality or from reasonably foreseeable misuse by persons'.
However functional insufficiency as term is currently not defined
in the terms and definition of the recent sate of the ISO/PAS 21448:2018 } (working title in the SOTIF context), \textbf{\emph{functional deficiency\index{functional deficiency}}}
and \textbf{\emph{\index{performance limitation}performance limitation}}
(I)\footnote{SOTIF ISO/PAS 21448 :2018: insufficiencies in the implementation of
the intended functionality}. The proposed terms are partly misleading as they might be understood
as related only to a sub-part of the actual problem, e.g. \emph{functional
insufficiency} might be interpreted as related only to the effect
of the deductive gap on the functional level (neglecting the remaining
horicontal layers illustrated in section \ref{subsec:Basic-concepts-devApproaches}).

The \textbf{\emph{validation challenge}} is closely related to the
many necessary deductions applied during development: for every step,
possibly leading to deductive gaps, 
\begin{itemize}
\item all underlying assumptions need to be made explicit
\item per assumption, evidence needs to be provided for its validity 
\end{itemize}
\textbf{\emph{Evidence}} is generated from meeting of expectations.
Hypotheses need to be formulated and their legitimacy needs to be
demonstrated, e.g. by formal approaches, simulation and real world
observation. Due to the complex open world problem, hypothesis checking
based on the developed model of reality $\bar{T}$ needs to be done
carefully, as applying the same approximations in derivation and interpretation
of hypothesis checks might lead to a self-fulfilling prophecy and
hence to a wrongly passed test, providing a \emph{misleading argument}
for the validity of the deduction (see the following section for a
detailed discussion).

\subsection{Development, iterative deduction and redesign \label{subsec:Development,-iterative-deduction}}

Figure \ref{fig:Triangle3new}b illustrates a typical situation during
development after $i$ steps of deduction, achieved by any approach
(layered design, V-model guided etc.). The deduction towards $\bar{T}_{i}$
(bold arrow 1) reduces the \emph{\ensuremath{\infty}\textendash complex}
and \emph{\ensuremath{\infty}\textendash voluminous} space of possibilities
to a more or less abstract (implying more or less implicit), however
finite \emph{representation} indicated by the dashed line. The related
explicit complexity is much lower than that of the (actually practically
not determinable) \emph{perfectly valid} $\tilde{T}$ - indicated
by the simple geometry of $\bar{T}$. In a nutshell, the development
process can be understood as iterative approaching a good enough (i.e.
\emph{valid}) approximation of a \emph{perfectly valid} $\tilde{T}$,
without knowing $\tilde{T}$ explicitly (see also the conception of
validity in section \ref{subsec:Conception-of-validity} and the discussion
related to $cond1$ in section \ref{subsec:Formalized-algorithm-for}
). 

As discussed in section \ref{subsec:Model-of-reality}, the deduced
triad $\bar{T}$ might not necessarily fully correctly map the true
characteristics $T^{\dagger}$. This deviation is depicted as deviation
of the dash-dot-dagger line from the dashed line for parts of the
model $\bar{T}$ in the lower part of figures \ref{fig:Triangle3new}b
-f. The validity of the developed system however is only related to
the deviation between the effective ($T^{\dagger}$) and the necessary
($\tilde{T}$). Due to this deviation between $T^{\dagger}$ and $\tilde{T}$,
four different source types of \emph{loss} might be inherent to the
resulting system, namely \emph{loss} (small bolts) or \emph{unacceptable
loss} (striped large bolts), each related to either unnecessarily
included subspace (e.g. unacceptable loss included in upper left of
figure \ref{fig:Triangle3new}b) or neglected relevant subspace (right
hand side of figure \ref{fig:Triangle3new}b), relative to $\tilde{T}$.

An important aspect of any reasonable development approach is encapsulation,
which means that a further deduction from $\bar{T}_{i}$ is based
on $\bar{T}_{i}$ solely. This however implies that any further deduction
$\bar{T}_{i}\downarrow\bar{T}_{i+1}$, as illustrated in figure \ref{fig:Triangle3new}
(bold arrow 2), can not reintroduce an already neglected subspace.
Therefore, an already inherent risk for the manifestation of loss
due to neglected relevant subspace can not be cured by further deduction
from $\bar{T}_{i}$. Only a redesign on (at least) level $i$, related
to a deduction $\bar{T}_{i-1}\downarrow\bar{T}_{j}$ (bold arrow 3),
might solve the missing relevant subspace issue as show in figure
\ref{fig:Triangle3new}d. The necessary level for redesign corresponds
to the level at which the relevant subspace got lost during deduction.
After redesigning, deduction can be continued, as illustrated by $\bar{T}_{j}\downarrow\bar{T}_{j+1}$
(bold arrow 4). Finally, a valid result might be achieved (i.e. a
good enough approximation of $\tilde{T}$ such, that no \emph{unacceptable
loss} might occur from the operation of the system. This is related
to the \emph{development goal}, discussed in the following section. 

In conclusion, this makes clear that validation aspects are present
on every layer of abstraction, more specifically, validation is strongly
related to every step of \emph{deduction}. These \emph{deductions}
occur frequently, also within an encapsulated coarse \emph{horizontal
layer} (such as e.g. system, functional and implementation layer).
Per \emph{deduction}, a sound validation argumentation is necessary,
to prevent major redesign expenditures later on. Aspects missed in
early phases of development are particularly problematic. Validation
therefore needs to be addressed seriously, right from the beginning
of development.

\subsection{The development goal\label{subsec:The-development-goal}}

From the previous sections, it becomes clear that complexity and interdependence
are the root causes for the manifestation of \emph{loss.} With respect
to complexity, one can differentiate two aspects, namely that related
to the\emph{ hyper-volume} of the solution space and, on the other
hand, the characteristics of the \emph{decision boundary} enclosing
the \emph{hyper-volume}. Starting from a \emph{high level goal}, the
complexity is mainly implicit and related to the possibly infinite
volume of the principal solution space. The development process maps
this \emph{\ensuremath{\infty}\textendash complex} starting point
onto a finite \emph{representation}, which has initially more abstract
(implicit) shares and is continuously driven towards a more explicit
approximation of the \emph{perfectly valid} solution. Thereby, the
considered solution space volume all in all is reduced. However, the
explicit complexity (related to the \emph{decision boundary} enclosing
this subspace) increases. Even though explicit, this complexity however
also contributes to the possible deviation between the mental/explicit\emph{
model} the developer has (the expected characteristics) and the true
effective characteristics of a built system (i.e. $T^{\dagger}-\bar{T}=\Delta_{model-real}(complexity)$). 

The \textbf{\emph{development goal}} therefore is not a fully explicit
expression of the \emph{perfectly valid} solution ($\tilde{T}$),
but rather an optimum between the implicit and explicit complexity
such, that the possible manifestation of \emph{loss} is reduced on
a \emph{robust basis}. More specifically, the system shall be free
from \emph{unacceptable loss}. With \textbf{\emph{robust basis}},
we refer to the fact that, for a complex system in an open context,
a formal proof of the validity of the underlying assumptions is impossible.
Robustness against deviation from what is expected (based on $\bar{T}$)
and effective operation of the system ($T^{\dagger}$) needs to be
designed into the system. In addition, validity of\emph{ assumptions
}and related validation argumentation needs to be continuously checked
post release. This is referred to as \textbf{\emph{ongoing validation}}\index{ongoing validation}. 

In conclusion, the iterative \emph{deduction} applied during development
- e.g. working through the \emph{horizontal layers} of abstraction
from the \emph{high level goal} to the final build product - serves
four basic purposes: 
\begin{enumerate}
\item deducting \emph{valid systems} at the given layer of abstraction,
preventing major design iterations later 
\item iterative reduction of implicit complexity 
\item keeping the related explicit complexity manageable 
\item providing metrics for hypothesis checking and evidence generation
on every layer for pre-release validation and post-release checking
and maintaining of validity. We refer to the derived metrics at the
heart of ongoing validation as \textbf{\emph{\index{assumption monitors}assumption
monitors.}}
\end{enumerate}

\section{Problem formalization\label{sec:Problem-formalization}}

\subsection{Full partition of solution space\label{subsec:Full-partition-of}}

\begin{figure}[H]
\begin{centering}
\includegraphics[viewport=100bp 100bp 865bp 510bp,clip,width=0.9\columnwidth]{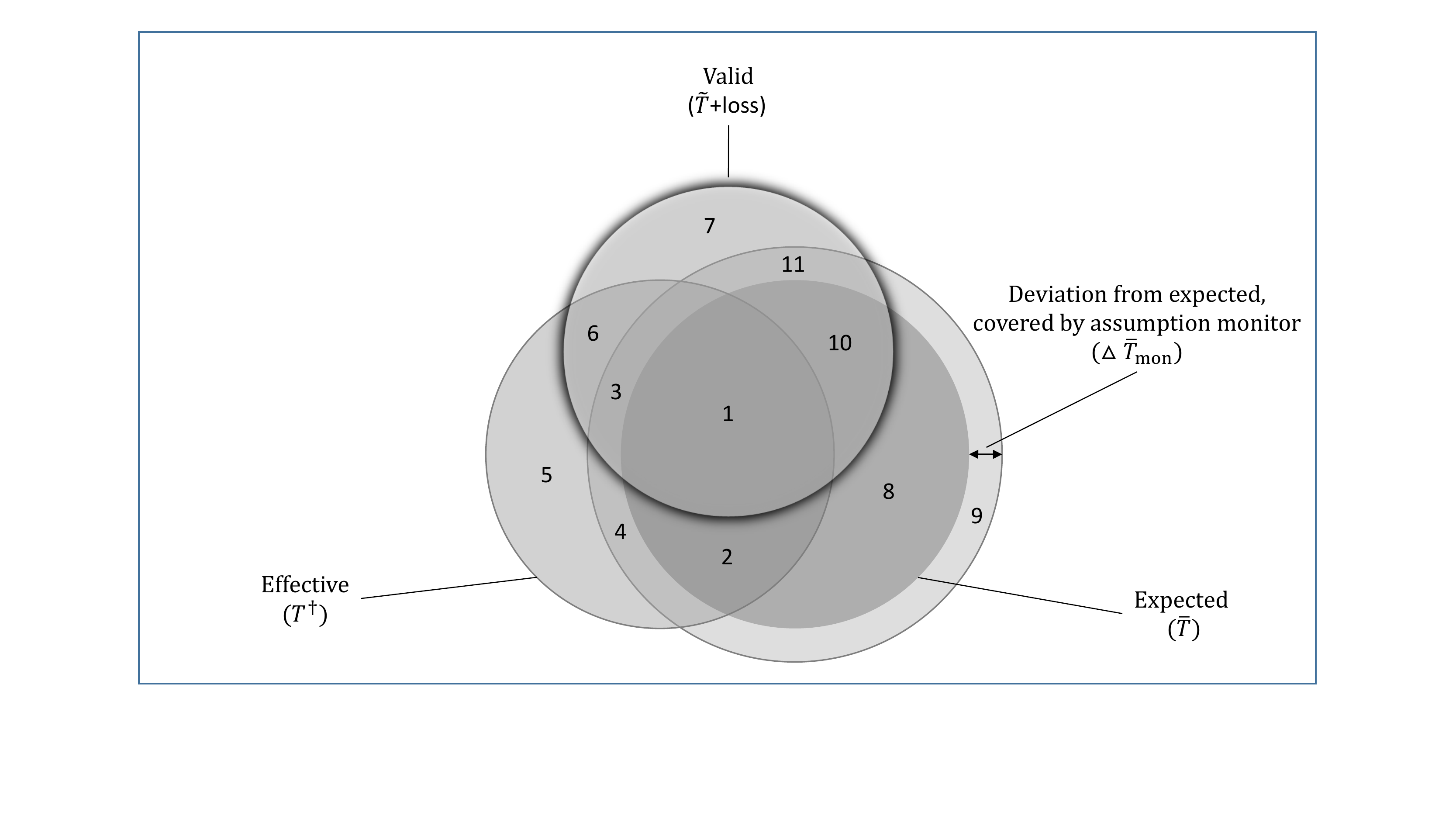}
\par\end{centering}
\caption{\label{fig:vennFull}The relation between the development goal (i.e.
a valid solution, see section \protect\ref{subsec:The-development-goal}),
the expected ($\bar{T}$) and the effective ($T^{\dagger}$, see section
\protect\ref{subsec:Model-of-reality}), as well as the role of assumption
monitoring (\emph{ongoing validation}) with referenced subareas, explained
in table \protect\ref{tab:vennArea}.}
\end{figure}

The relation among the development goal (i.e. a valid solution, see
section \ref{subsec:The-development-goal}), the expected ($\bar{T}$)
and the effective ($T^{\dagger}$, see section \ref{subsec:Model-of-reality}),
as well as the role of assumption monitoring (discussed in section
\ref{subsec:Deductive-gap,-validation}, \ref{subsec:Development,-iterative-deduction}
and \ref{subsec:The-development-goal}) is visualized in figure \ref{fig:vennFull}.

We apply the following notation: \textbf{labeled edges} refer to the
complete related circle area; we have:
\begin{itemize}
\item Valid solutions. I.e. perfectly valid $\tilde{T}$ plus the set of
solutions bearing possible loss which is below unacceptable loss,
indicated by the black (diffuse) edge. This is the development goal
(see section \ref{subsec:The-development-goal}). The diffuse style
relates to the fuzziness of the discrimination between loss and unacceptable
loss (it depends on attributes like type of cause of \emph{loss} and
societal factors as discussed in section \ref{subsec:Conception-of-validity}).
\item The expected $\bar{T}$ on the lower right of the diagram. It relates
to all the mental and explicit models (see section \ref{subsec:Model-of-reality}).
\item The effective ($T^{\dagger}$) on the lower left (see section \ref{subsec:Model-of-reality}).
\end{itemize}
\textbf{Labeled areas} refer to the related circular ring area, here
the area labeled by $\Delta\bar{T}_{mon}$referring to any deviation
from the expected ($\bar{T}$), which is covered by \emph{assumption
monitor}s ($\Delta\bar{T}_{mon}$).

As discussed in connection with the deductive gap (section \ref{subsec:Deductive-gap,-validation}),
iterative deduction and redesign (section \ref{subsec:Development,-iterative-deduction})
and ongoing validation (section \ref{subsec:The-development-goal}),
assumptions and related hypothesis checking play an important role
for validation. The assumption-related measures (like monitoring)
are an inevitable add-on to the models ($\bar{T}$), intentionally
extrinsic to $\bar{T}$, in order to overcome the \emph{misleading
argument} problematic. Intentionally extrinsic means that we aim for
assumption handling not prone to the same limitations than the model.
Making assumptions explicit and monitoring them prevents the \emph{misleading
argument} and allows for incidents\footnote{With incident we refer to any event leading to a monitor indicating
a deviation. Sound monitoring allows identifying such deviations,
even before any loss manifests.} triggering specific rework for improvement during development and
post-release ongoing validation. The aim is to maximize this area
such, that the effective ($T^{\dagger}$) is fully covered. 

\textbf{Numbers} refer to the subareas of circular intersections (the
area having one effective color from overlays). Area 1, for example,
indicates the subset of solutions, which effectively are as expected
and valid (overlay of $T^{\dagger}$, $\bar{T}$ and $\tilde{T}+$loss).
This, in fact, is the area that we aim to maximize.

Table \ref{tab:vennArea} explains all subareas referenced in figure
\ref{fig:vennFull}.

\begin{table}
\centering{}%
\begin{tabular}{|c|>{\raggedright}p{0.92\textwidth}|}
\hline 
\textsf{\textbf{\footnotesize{}Id}} &
\textsf{\textbf{\footnotesize{}Description}}\tabularnewline
\hline 
\hline 
\textsf{\footnotesize{}1} &
\textsf{\footnotesize{}The space of solutions being effectively as
expected and valid. This is the area we aim to maximize.}\tabularnewline
\hline 
\textsf{\footnotesize{}2} &
\textsf{\footnotesize{}The space of solutions being effectively as
expected, however invalid, e.g. due to invalid assumption. Due to
the congruence of the expected ($\bar{T}$) and the effective ($T^{\dagger}$),
the deviation can be found via dedicated verification and validation
(v\&v) efforts based on $\bar{T}$.}\tabularnewline
\hline 
\textsf{\footnotesize{}3} &
\textsf{\footnotesize{}The effective deviates from the expected and
is (by chance) valid. Due to the deviation between the effective ($T^{\dagger}$)
and the expected ($\bar{T}$), the deviations are not accessible via
dedicated v\&v efforts based on $\bar{T}$. However, the deviation
is accessible via }\textsf{\emph{\footnotesize{}assumption monitoring}}\textsf{\footnotesize{}.}\tabularnewline
\hline 
4 &
\textsf{\footnotesize{}The effective deviates from the expected and
is invalid. Due to the deviation between the effective ($T^{\dagger}$)
and the expected ($\bar{T}$), the deviations are not accessible via
dedicated v\&v efforts based on $\bar{T}$. However, the deviation
is accessible via}\textsf{\emph{\footnotesize{} assumption monitoring}}\textsf{\footnotesize{}.
The monitors allow to identify already smaller incidents. The assumption
related efforts allow to trigger specific rework for improvement.}\tabularnewline
\hline 
\textsf{\footnotesize{}5} &
\textsf{\footnotesize{}The effective deviates from the expected and
is invalid. Due to the deviation between the effective ($T^{\dagger}$)
and the expected ($\bar{T}$), the deviations are not accessible via
dedicated v\&v efforts based on $\bar{T}$. In addition, the deviation
is not accessible via }\textsf{\emph{\footnotesize{}assumption monitoring}}\textsf{\footnotesize{}.
Invalidity can only be discovered by manifestation of unacceptable
loss; this however can not be explained based on the present models.
Such an event triggers major rework.}\tabularnewline
\hline 
\textsf{\footnotesize{}6} &
\textsf{\footnotesize{}The effective deviates from the expected and
is (by chance) valid. Due to the deviation between the effective ($T^{\dagger}$)
and the expected ($\bar{T}$), the deviations are not accessible via
dedicated v\&v efforts based on $\bar{T}$. No }\textsf{\emph{\footnotesize{}assumption
monitoring.}}\textsf{\footnotesize{} This is the region related to
the }\textsf{\emph{\footnotesize{}misleading argument}}\textsf{\footnotesize{}
problematic as the wrongful expectation might apparently be supported
by dedicated v\&v efforts based on $\bar{T}$.}\tabularnewline
\hline 
\textsf{\footnotesize{}7} &
\textsf{\footnotesize{}Valid design space not used. This might be
inconvenient as in principle available design space is not used for
optimization of product properties, or even be problematic (from a
validation point of vie), as implicitly expected properties of the
product are not achieved. }\tabularnewline
\hline 
\textsf{\footnotesize{}8} &
\textsf{\footnotesize{}Invalid model, however never effectively reached
part of the solution space.}\tabularnewline
\hline 
\textsf{\footnotesize{}9} &
\textsf{\footnotesize{}In principle monitored deviation from the expected
(however invalid). The effective deviates from the expected and never
reaches this part of the solution space $\rightarrow$never effectively
triggered }\textsf{\emph{\footnotesize{}assumption monitor}}\textsf{\footnotesize{}.}\tabularnewline
\hline 
\textsf{\footnotesize{}10} &
\textsf{\footnotesize{}The expected would be valid, however the effective
deviates and never reaches this part of the solution space. }\tabularnewline
\hline 
\textsf{\footnotesize{}11} &
\textsf{\footnotesize{}In principle monitored deviation from the expected
(valid), however the effective deviates from the expected and never
reaches this part of the solution space $\rightarrow$never effectively
triggered }\textsf{\emph{\footnotesize{}assumption monitor}}\textsf{\footnotesize{}.}\tabularnewline
\hline 
\end{tabular}\caption{\label{tab:vennArea}Description of the subareas of the intersection
between the development goal (i.e. a valid solution, see section \protect\ref{subsec:The-development-goal}),
the expected ($\bar{T}$) and the effective ($T^{\dagger}$, see section
\protect\ref{subsec:Model-of-reality}), as well as the role of assumption
monitoring (\emph{ongoing validation}) referenced in figure \protect\ref{fig:vennFull}. }
\end{table}
The area 6 relating to solutions that are prone to the \emph{misleading
argument} problematic is especially critical in the region close to
the fuzzy boundary between loss and unacceptable loss. Even when conscientiously
following the societal discussion and in principle being prepared
to respond to shifts, a (by chance) valid effective solution of type
6 may become unnoticed invalid (at least until an unacceptable loss
occurs), as the model $\bar{T}$ spuriously relates the solution to
a different solution space which is not affected by the shift. 

In conclusion, with respect to the areas illustrated in figure \ref{fig:vennFull},
a development process should address the following aspects:
\begin{enumerate}
\item Maximize area 1. This is the ultimate goal, which needs to be supported
in initial design, during development and post release in the sense
of continuous improvement. All the following aspects contribute to
this.
\item Provide support for identifying the space of valid solutions (circle
$\tilde{T}+$loss), especially addressing the diffuse boundary between
loss and unacceptable loss (black diffuse edge). 
\item Provide support for derivation of valid models (circle $\bar{T}$),
necessarily accompanied by support in identifying and validation of
underlying assumptions, and maximizing circular ring $\Delta\bar{T}_{mon}$.
\item Provide support for robust design: large however expected deviations
from the nominal should be taken into account in development. I.e.
don't design a system mainly operating close to the fuzzy boundary. 
\item Derive sound dedicated v\&v contributions based on $\bar{T}$ and
underlying assumptions (addressing area 2).
\item Be prepared to the unexpected. I.e. maximize $\Delta\bar{T}_{mon}$
(in 3.) and establish the possibility to identify already minor incidents
and trigger specific rework for improvement. This is beyond robustness
and related to what Taleb refers to as \textbf{\emph{anti-fragility}}
- a system under stress needs to survive and get better \cite{Taleb2012,Taleb2013,Taleb2014a}.
Especially maximize area 3 and 4, thereby minimizing 5 and 6. 
\item Support design freedom by using the available design space (i.e. minimize
7). In a nutshell, the process should allow addressing the above mentioned
aspects and identifying design drivers, while preventing a too early
focusing on certain implementations (which would effectively reduce
the available solution space). As an example, a system level hazard
and risk analysis allows identifying design drivers in an early stage
of development.
\end{enumerate}

\subsection{Formalized algorithm for iterative development and validation\label{subsec:Formalized-algorithm-for}}

\subsubsection*{Initial remark}

In this section, formalized algorithms (and the necessary ingredients)
for the development and iterative validation of complex systems, operating
in open contexts, are provided. Due to the complexity of the validation
issue, this formalization is regarded necessary in order to ensure
that none of the many subtle details are neglected in development
process design. 

The reader being mainly interested in the broad mindset might skip
the formalized algorithms and head on for section \ref{sec:contribAll},
after ensuring he is aware of the fact that $cond1$, given below,
can practically only be enforced indirectly (via eq. (\ref{eq:rep})
and $cond2$). All other basic aspects have already been discussed
in the foregoing sections, and a high level overview of a practical
process is given in section \ref{sec:contribAll}.

Figure \ref{fig:algoFlow} provides a high level overview of the complex
interdependence and mutually recursive flow through the algorithms.

\subsubsection*{Definitions and ingredients}

A \textbf{\emph{representation}} is a set of statements representing
the possibly $\infty$-complex reality. A \emph{representation} might
be built from a wide range of elements, e.g. abstract textual statements
or even an expansion of reality in an explicit basis sets in the mathematical
sense. \\

Any given \emph{representation} therefore has a certain \emph{abstraction
level} whereas higher abstraction levels are related to more implicity,
leaving room e.g. for deviating interpretation (i.e. alternative concretizations).
\\

Transformation of \emph{representation} to another \emph{representation}
is indicated by $\bar{T}_{i}\downarrow\bar{T}_{i+1}$\\

The actual goal during development is to achieve a sufficiently explicit
\emph{representation} of the relevant part of reality (e.g. the aspects
of the \emph{validation triangle}) in order to be able to argue about
the validity of the achieved solution (see section \ref{subsec:The-development-goal}
for details). More formally, for complex systems operating in an open
context, the condition is

\begin{equation}
cond1:\,(\bar{T}\stackrel{suff\land robust}{\approx}T^{\dagger})\land(T^{\dagger}\in\{(\tilde{T}+loss)_{robust}\})\label{eq:cond1}
\end{equation}

I.e. the model of reality $\bar{T}$ needs to sufficiently well approximate
the effective $T^{\dagger}$, this approximation needs to be robust
against stress, and the effective needs to be in the subset of valid
solutions $\{(\tilde{T}+loss)_{robust}\}$ which robustly stay valid
even under stress. In addition, possible deviations need to be recognized
for antifragility (i.e. a system under stress needs to get aware of
this and means to survive and get better need to be prepared), see
\emph{assumption monitors} in the following.\\

Neither the effective $T^{\dagger}$, nor the valid $\tilde{T}+loss$
are directly accessible (as discussed in the foregoing sections).
Therefore, achievement of $cond1$ can not directly be argued / examined.
Instead, development and argumentation about the system is based on
\emph{representations} of reality, which consists of models of reality
$\bar{T}$ related to the aspects of the validation triangle, accompanied
by further elements as introduced in the following. $Cond1$ is then
indirectly enforced via the complex, however accessible condition
$cond2$ given below.\\

The \emph{high level goal} is the most abstract \emph{representation}
of the implicit, $\infty$-complex product idea (which initially is
free from \emph{representation}). Following system-view based approaches
(e.g. STAMP \cite{Leveson2011}), a sufficiently complete set of high
level \emph{unacceptable losses} can be derived from the high level
goal. Based on this, a set of high level \textbf{\emph{constraints}}
$\{Cr\}$ can be formulated, which constrain the system away from
states which might, under certain conditions, lead to the manifestation
of the determined \emph{unacceptable losses}.\\

The initial formation of \emph{representation} of the implicit and
$\infty$-complex reality related to the product idea, just as every
following transformation of \emph{representation}, usually involves
a set of assumptions $\{A\}$. These underlying assumptions need to
be elicitated by a conscientious process and the sufficiently completeness
and validity needs to be argued (see evidence-based argumentation
below, which is required also for other elements, such as $\{Cr\}$).\\

The basis for the indirect approach to condition $cond1$, in a nutshell,
is formed by two main contributions, namely the arguable validity
of underlying\emph{ assumptions} and the enforcement of (arguable
valid) \emph{constraints} $\{Cr\}.$ Each contribution per se can
never be perfect, but could in principle ensure the validity of the
system. This approach therefore forms a mutual reinforcing double
barrier against \emph{unacceptable losses}\footnote{Suppose all underlying assumptions would be valid. Then the model
of reality, reasoning about the system and its context and the derived
argumentation for safe operation would be valid. Hence manifestation
of unacceptable losses would be prevented. In the case some assumptions
might get invalidated, the valid enforcement of the constraints $\{Cr\}$
would still prevent the manifestation. }.\\

As discussed in the foregoing sections (namely sections \ref{subsec:Deductive-gap,-validation}
- \ref{subsec:Full-partition-of}), an arguably sufficiently complete
set of \emph{monitors} needs to be derived in order to implement robustness
and antifragility. The already introduced \emph{assumption monitor}s
($\{\Delta\bar{T}_{mon}\}$) need, as argued in the foregoing passages,
to be supplemented by \textbf{\emph{satisfaction monitors for constraints}}\emph{
$\{Cr_{mon}\}$} \\

Robustness and antifragility is achieved by derivation of \emph{recovery-}
and \emph{degradation strategies} in case a \emph{monitor} indicates
the system being close to or already in violation of the related requirement.
\textbf{\emph{Recovery}} thereby is related to a situation in which
normal operation can be re-achieved, whereas \textbf{\emph{degradation}}
relates to a degradation of the systems functionality in order to
prevent manifestation of unacceptable loss (e.g. transition to safe
stop).
\begin{align*}
\{recDegStrat\} & :=\{recovStrat\}\cup\{degStrat\}
\end{align*}
with assumptions- and constraints related contributions:
\begin{align*}
\{recovStrat\} & :=\{recovStrat^{^{\Delta\bar{T}_{mon}}}\}\cup\{recovStrat^{Cr_{mon}}\}\\
\{degStrat\} & :=\{degStrat^{^{\Delta\bar{T}_{mon}}}\}\cup\{degStrat^{Cr_{mon}}\}
\end{align*}

The sufficient \emph{representation} of the aspects of the validation
triangle $\bar{T}$, as well as the sufficient completeness of the
set of constraints $\{Cr\}$, elicitated assumptions $\{A\}$ and
derived assumption monitors $\{\Delta\bar{T}_{mon}\}$ need to be
argued, based on evidence. As discussed above, the argumentation needs
to address aspects such as application of appropriate processes (e.g.
for elicitation), sufficient completeness of sets (e.g. the set of
assumptions ${A}$) and aspects of robustness and antifragility. We
subsume all these aspects of argumentation in the operator $\vartriangleleft$
and write, e.g. for the set of evidence based arguments related to
constraints $(\{evArg^{Cr}\})$:
\[
\{evArg^{Cr}\}:=\{Cr\}\vartriangleleft\{evArg\}
\]

The complete set of evidence based arguments is denoted as $\{evArg\}$,
with
\[
\{evArg\}:=\{evArg^{\bar{T}}\}\cup\{evArg^{Cr}\}\cup\{evArg^{A}\}\cup\{evArg^{\Delta\bar{T}_{mon}}\}\cup\{evArg^{Cr_{mon}}\}\cup\{evArg^{recDegStrat}\}.
\]

The argumentation is relative to the \emph{representation} from which
it has been deducted, or reality in the case of the highest level
deduction\footnote{Argumentation relative to the \emph{representation} deduced from follows
the well established divide and conquer approach. It is not manageable
to argue about every transformation relative to the full implicit
reality. Instead, each \emph{representation} is argued to be valid
and following transformations can then be argued relative to this
(arguably valid) reference. }.\\

There needs to be defined a set of quality criteria $\{Qc\}$ for
the evidence based argumentation allowing to rate the state of the
argumentation (i.e. insufficient or sufficient). As discussed in section
\ref{subsec:Development,-iterative-deduction}, validity of a certain
state of development should be ensured (i.e. $\{evArg\}$ complies
with $\{Qc\}$) before applying further transformations. Note that
the quality criteria for the combination of the diverse and fragmented
evidence contributions to an argumentation is closely related to the
societal expectations discussed in the foregoing sections. Derivation
of $\{Qc\}$ therefore is a complex task.\\

For preconditions, we use the notation $pre$.

\subsubsection*{Indirectly enforcing validity}

To sum up, a \emph{representation} at level $i$ is given by
\begin{align}
\{Rep_{i}\}: & \{\bar{T}_{i}\}\cup\{Cr_{i}\}\cup\{A_{i}\}\cup\{\Delta\bar{T}_{mon,i}\}\cup\{Cr_{mon,i}\}\cup\{recDegStrat_{i}\}\cup\{evArg_{i}\}\cup\{Qc_{i}\}\label{eq:rep}
\end{align}
and condition $cond1$ is enforced by: 
\begin{align}
cond2: & \forall i\in\{0\ldots|\{Rep\}|\}\exists\{Rep_{i}\}|\{evArg_{i}\}complies\,with\{Qc_{i}\}\label{eq:cond2}
\end{align}
which can be stated as: \emph{for all underlying representations,
the representations are complete according to (\ref{eq:rep}) and
the evidence based argumentation complies with the given quality criteria}.\\

\subsubsection*{Overview of the flow through the algorithms}

\begin{figure}[H]
\begin{centering}
\includegraphics[width=1\columnwidth]{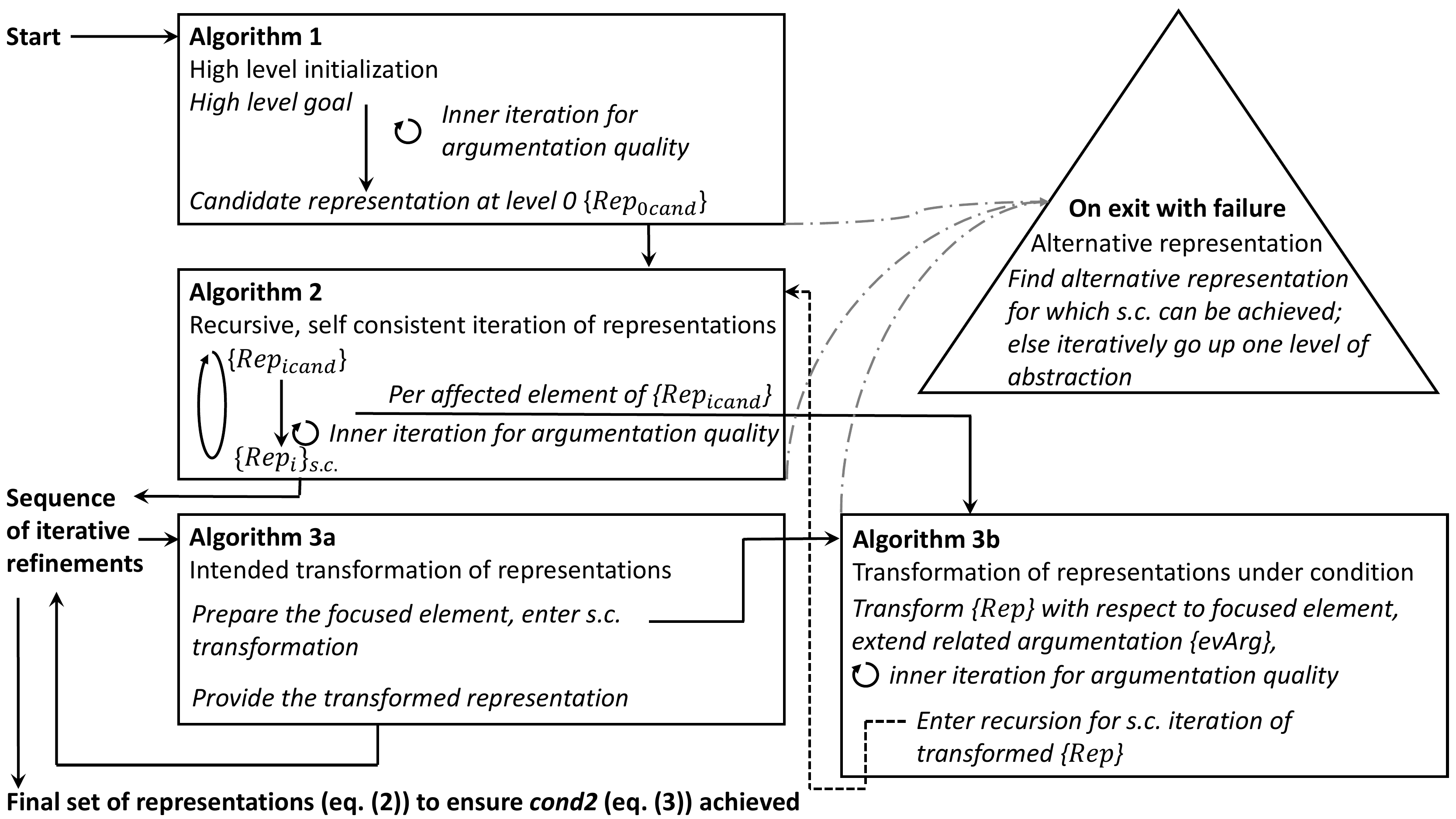}
\par\end{centering}
\caption{\label{fig:algoFlow}high level overview of the complex interdependence
and mutually recursive flow through the algorithms.}
\end{figure}

Figure \ref{fig:algoFlow} provides a high level overview of the complex
interdependence and mutually recursive flow through the algorithms. 

After successful high level initialization (providing a self consistent
highest level representation $\{Rep_{0}\}s.c$), a sequence of iterative,
intended refinements (subsequent application of algo3a) is undertaken
such that an appropriate set of representations (across multiple level
of abstractions, according to eq. ($\ref{eq:rep}$)) is achieved and
$cond2$ (eq. (\ref{eq:cond2})) is satisfied. Algorithm 2 thereby
iteratively drives a representation to self-consistency, which may
require transformations of the representation (algorithm 3b). Vice
versa, a transformed representation needs, just as well, be brought
to self-consistency. Therefore, algorithm 2 and 3b are mutual dependent,
with possibly several alternating recursions becoming necessary. 

\subsubsection*{Algorithm 1 - high level initialization}

\rule[0.5ex]{1\columnwidth}{1pt}

Algo1: $($initial product idea$)\longrightarrow\{Rep_{0}\}s.c.$

\rule[0.5ex]{1\columnwidth}{0.5pt}

\begin{enumerate}
\item start from initial product idea, derive the \emph{high level goal} 
\item establish general quality criteria for the underlying processes and
validity of individual aspects $\{Qc\}$ 
\item from the \emph{high level goal} derive a candidate representation
$\{Rep_{0cand}\}:$
\begin{enumerate}
\item carry out initial design domain and functional analysis, set up open
model $\{\bar{T}_{0}\}=\{\bar{C}_{0}\}\cup\{\bar{P}_{0}\}\cup\{\bar{R}_{0}\}$
\item apply conscientious process to elicitate related $\{A_{0}\}$
\item carry out analysis of high level \emph{unacceptable losses}, derive
$\{Cr_{0}\}$, possibly extending $\{A_{0}\}$, note this in $\{exA_{0}\}$
\item derive $\{\Delta\bar{T}_{mon,0}\}$ from $\{A_{0}\}$, possibly extending
$\{A_{0}\}$, note this in $\{exA_{0}\}$
\item derive $\{Cr_{mon,0}\}$ from $\{Cr_{0}\}$, possibly extending $\{A_{0}\}$,
note this in $\{exA_{0}\}$
\item derive $\{recovStrat_{0}^{^{\Delta\bar{T}_{mon}}}\}$ and $\{degStrat_{0}^{^{\Delta\bar{T}_{mon}}}\}$,
possibly extending $\{A_{0}\}$, note this in $\{exA_{0}\}$
\item derive $\{recovStrat_{0}^{Cr_{mon}}\}$ and $\{degStrat_{0}^{Crmon}\}$,
possibly extending $\{A_{0}\}$, note this in $\{exA_{0}\}$
\item refine $\{Qc\}$ if necessary:$\{Qc\}\downarrow\{Qc_{0}\}$, possibly
extending $\{A_{0}\}$, note this in $\{exA_{0}\}$
\item compile evidence based argumentation $\{evArg_{0}\}$, possibly extending
$\{A_{0}\}$, note this in $\{exA_{0}\}$, related to: 
\begin{enumerate}
\item derivation process and individual validity of the aspects of the validation
triangle
\begin{eqnarray*}
\{\bar{T}_{0}\}\vartriangleleft\{evArg\} & = & \{\bar{C}_{0}\}\vartriangleleft\{evArg\}|_{\{\bar{P}_{0}\},\{\bar{R}_{0}\}}\\
 &  & \cup\{\bar{P}_{0}\}\vartriangleleft\{evArg\}|_{\{\bar{C}_{0}\},\{\bar{R}_{0}\}}\\
 &  & \cup\{\bar{R}_{0}\}\vartriangleleft\{evArg\}|_{\{\bar{C}_{0}\},\{\bar{P}_{0}\}}
\end{eqnarray*}
\item elicitation process and individual validity of underlying assumptions
$\{A_{0}\}\vartriangleleft\{evArg\}$
\item derivation process and individual validity of assumption monitors
$\{\Delta\bar{T}_{mon,0}\}\vartriangleleft\{evArg\}$
\item derivation process and individual validity of constraints $\{Cr_{0}\}\vartriangleleft\{evArg\}$
\item derivation process and individual validity of satisfaction monitors
for constraints $\{Cr_{mon,0}\}$
\item derivation process and individual validity of $\{recDegStrat_{0}\}\vartriangleleft\{evArg\}$ 
\end{enumerate}
\item iteratively enhance argumentation quality until $\{evArg_{0}\}complies\,with\{Qc_{0}\}$;
if iteration fails, exit with failure
\item enter self-consistency loop (algo2) based on the candidate \emph{representation}
for the highest abstraction level $\{Rep_{0cand}\}$ and $\{exA_{0}\}$,
which gives:\\
$\{Rep_{0}\}s.c.=$Algo2$(\{Rep_{0cand}\},\{exA_{0}\})$ 
\end{enumerate}
\end{enumerate}

\rule[0.5ex]{1\columnwidth}{0.5pt}

\subsubsection*{Algorithm 2 - recursive, self consistent iteration of \emph{representations}}

As it becomes clear from the listing of the high level initialization
algorithm, the elements of the \emph{representation} are mutually
dependent. Specifically, the list of $\{exA_{i}\}$ grows from algo1
3a) to algo1 3k), with the prior items not being consistent with the
complete list of assumptions after a singular pass. Therefore, all
elements of the candidate \emph{representation} need to be checked
and possibly extended with respect to all elements in $\{exA_{i}\}$.
This can extend $\{Rep_{icand}\}$ and introduce a new set of $\{exA_{i}\}$.
Therefore, another cycle of checking the updated $\{Rep_{icand}\}$
against the updated $\{exA_{i}\}$ needs to be started until $\{exA_{i}\}=\emptyset$,
which implies self-consistency of the \emph{representation}: 
\begin{align*}
\{Rep_{icand}\}\longrightarrow\{Rep_{i}\}\,s.c.
\end{align*}

\rule[0.5ex]{1\columnwidth}{1pt}

Algo2: $(\{Rep_{icand}\},\{exA_{i}\})\longrightarrow\{Rep_{i}\}s.c.$

\rule[0.5ex]{1\columnwidth}{0.5pt}

\begin{enumerate}
\item set $\{Rep_{cand}\}=\{Rep_{icand}\}$
\item set $\{exA\}=\{exA_{i}\}$
\item set $\{exA_{i}\}=\emptyset$
\item For $j=1,\ldots,|\{exA\}|$:
\begin{enumerate}
\item check all elements of $\{\bar{T}_{icand}\}$ with respect to $exA^{j}\in\{exA\}$.
If affected, transform the element using algo3b such that the resulting
\emph{representation} is consistent with $exA^{j}$. I.e. 
\begin{enumerate}
\item set affected element as focused element for next transformation $\{f_{in}\}$
\item $\{Rep_{cand}\}=$Algo3b$(\{Rep_{cand}\},\{f_{in}\},exA^{j})$
\end{enumerate}
This possibly updates $\{Rep_{cand}\}$.
\item check and transform $\{A_{icand}\}$accordingly; this possibly extends
$\{exA_{i}\}$ (which was set empty in step algo2 3.)
\item check and transform $\{recDegStrat_{icand}\}$, $\{Cr_{icand}\}$,
$\{\Delta\bar{T}_{mon,icand}\}$, $\{Cr_{mon,icand}\}$, $\{Qc_{icand}\}$
and $\{evArg_{icand}\}$ accordingly, possibly extending $\{A_{cand}\}$
and $\{exA_{i}\}$
\item check and transform $\{evArg_{icand}\}$ accordingly (applying the
sequence alg1.3.i), possibly extending $\{A_{cand}\}$ and $\{exA_{i}\}$
\item iteratively enhance argumentation quality until $\{evArg_{cand}\}complies\,with\{Qc_{cand}\}$;
if iteration fails, exit with failure.\\
This sequence results in an updated $\{Rep_{cand}\}$ with respect
to $exA^{j}$ being the basis for $j+1$ and a possibly non-empty
$\{exA_{i}\}$.
\end{enumerate}
\item The above for loop results in an updated $\{Rep_{cand}\}$ with respect
to $\{exA\}$ being the basis for the next self consistency cycle
and a possibly non-empty $\{exA_{i}\}$.\\
If $\{exA_{i}\}=\emptyset$ return $\{Rep_{i}\}\,s.c.=$$\{Rep_{cand}\}$\\
else set $\{Rep_{icand}\}=\{Rep_{cand}\}$ and re-start algo2
\end{enumerate}

\rule[0.5ex]{1\columnwidth}{0.5pt}

\subsubsection*{Algorithm 3a - intended transformation of \emph{representations}}

\rule[0.5ex]{1\columnwidth}{1pt}

Algo3a: $\{Rep_{i}\}s.c.\longrightarrow\{Rep_{i+1}\}s.c.$

\rule[0.5ex]{1\columnwidth}{0.5pt}

\begin{enumerate}
\item $pre:\{evArg_{i}\}\,complies\,with\{Qc_{i}\}\land\{Rep_{i}\}\,s.c.$\\
From $\{\bar{T}_{i}\}=\{\bar{C}_{i}\}\cup\{\bar{P}_{i}\}\cup\{\bar{R}_{i}\}$
select one or several aspects as focused elements for next transformation:
\[
\{f_{i}\}\subseteq\{\bar{T}_{i}\}
\]
\item $\{Rep_{i+1}\}=$Algo3b$(\{Rep_{i}\},\{f_{i}\},\emptyset)$
\item return $\{Rep_{i+1}\}s.c.=$$\{Rep_{i+1}\}$
\end{enumerate}

\rule[0.5ex]{1\columnwidth}{0.5pt}

\subsubsection{Algorithm 3b - transformation of \emph{representations} under condition}

\rule[0.5ex]{1\columnwidth}{1pt}

Algo3b: $(\{Rep_{in}\},\{f_{in}\},\{exCond_{in}\})\longrightarrow\{Rep_{out}\}$

\rule[0.5ex]{1\columnwidth}{0.5pt}

\begin{enumerate}
\item set $\{exA_{i}\}=\emptyset$
\item For $j=1,\ldots,|\{f_{in}\}|$:
\begin{enumerate}
\item derive transformed \emph{representation} of focused element, all other
aspects held fixed, such that extra consistency conditions $\{exCond_{in}\}$
are fulfilled:
\[
f_{in}^{j}\downarrow\{f_{in+1}^{j}\}|\left((\{Rep_{in}\}\setminus f_{in}^{j}=const.)\land(\{Rep_{in}\}\setminus f_{in}^{j}\,consistend\,to\,\{exCond_{i}\})\right)
\]
The set notation indicates the fact that the transformation
of one element $f_{i}^{j}$ might lead to a set of elements $\{f_{i+1}^{j}\}$.
If impossible, exit with failure.
\item set $\{Rep_{out}\}=(\{Rep_{in}\}\setminus f_{in}^{j})\cup\{f_{in+1}^{^{j}}\}$
\item extend $\{Rep_{out}\}$ by 
\begin{enumerate}
\item apply conscientious process to elicitate transformation related $\{A\}$
\item derive $\{recDegStrat\}$ , possibly extending $\{A\}$, note this
in $\{exA\}$
\item derive $\{\Delta\bar{T}_{mon}\}$ from $\{A\}$, possibly extending
$\{A\}$, note this in $\{exA\}$
\item extend evidence based argumentation $\{evArg\}$, applying the sequence
alg1 3.i, possibly extending $\{A\}$, note this in $\{exA\}$ following 
\item iteratively enhance argumentation quality until $\{evArg\}complies\,with\{Qc\}$;
if iteration fails, exit with failure.
\end{enumerate}
\item This results in an updated $\{Rep_{out}\}$ 
\item If $\{exA_{i}\}=\emptyset$ exit\\
else 
\begin{enumerate}
\item set $\{Rep_{icand}\}=\{Rep_{out}\}$
\item set $\{exA_{i}\}=\{exA\}$
\item $\{Rep_{out}\}=$Algo2$(\{Rep_{out}\},\{exA\})$
\item return $\{Rep_{out}\}$
\end{enumerate}
\end{enumerate}
\end{enumerate}

\rule[0.5ex]{1\columnwidth}{0.5pt}

\subsubsection*{On exit with failure}

Exit with failure might happen at algo1 3.j, algo2 4.e, algo3b 2.a
and algo3b 2.c.v. The reason is that the underlying \emph{representation}
can not be brought to consistent validity, which means that an alternative
transformation on the current level needs to be found. If this is
still not possible, one needs to iteratively step up one layer at
a time ($\uparrow\{Rep_{i-1})$) and try to find a working transformation
on that layer. Finding an alternative transformation on a layer close
to the failed layer (most preferentially on the same layer) is beneficial,
as stepping up one layer in each case renders already invested validation
related results from $\{Rep_{j-1})\downarrow\{Rep_{j})$ unused. Typically,
for alternative transformations, only a part of these validation results
can be reused. 

\section{Development process, contributions and standards\label{sec:contribAll}}

\subsection{Overview\label{subsec:Overview-proc}}

Based on the discussion in the foregoing chapters, we briefly sketch
a \emph{holistic} development process which we refer to as systematic,
system view based approach to validation, in short \textbf{\emph{sys$^{2}$val}}.
A detailed presentation however is out of the scope of this paper
and might be provided later on. A discussion of published approaches
by others, based on our presentation of the validation challenge,
will be published later on.

\textbf{\emph{Holistic}} in this context means that the process needs
to address the whole process (all aspects of the listing in section
\ref{subsec:Full-partition-of}), from initial framing of intention
(\emph{high level goal}) over design and implementation to post-release
operation, supporting evidence generation from fragmentary, manifold
sources (such as simulation, real world driving, etc.), \emph{ongoing
validation} and continuous improvement.

\emph{\small{}Sys$^{2}$val}\emph{,} just as our general mindset,
is strongly influenced by the work of Rasmussen and Leveson \cite{Rasmussen1983,Rasmussen1994,Leveson2011}.
We build on this strong basis and extend it to enhance traceability,
completeness- \& validation argumentation and applicability for complex
systems operating in open contexts. 

On the highest level, a holistic process should support framing of
the intention (i.e. high level goal definition), taking in to account
all relevant aspects (such as system safety, safety of the intended
functionality, functional safety, security,product liability, etc.)
listed in section \ref{subsec:Conception-of-validity}.

Next, just as in Leveson's approach, the 'what and where should it
be done' (high level purpose and operational design domain analysis)
is initially separated from the aspect of 'what should not happen
in order to not compromise the high level goal'. This analysis is
effect based (i.e. based on high level accidents / unacceptable losses).
It necessarily is - as far as possible - independent from specific
solution approaches and details of implementation. The aim is to provide
general applicable constraints - rated by a related risk level, to
the following design and implementation without reducing the accessible
solution space right from the start (maximize area 7, point 7 in the
listing at end of section\ref{subsec:Full-partition-of} ). On the
high level of abstraction, an arguable complete set of hazardous system
states and related constraints can be determined, which is easily
comprehensible. The derived constraints contribute to the practically
only indirect enforcing of $cond1$ via eq. ($\ref{eq:rep}$) and
$cond2$, as discussed in section \ref{subsec:Formalized-algorithm-for}. 

One of the basic aspects of \emph{Sys$^{2}$val} is to apply what
we refer to as \textbf{\emph{open model approach}}, which is a model
approach coping with the complexity of open contexts, allowing for
iterative return to the model building, validation and refinement
without the need to fully explicit modeling of e.g. the operational
design domain (which would be impossible in an open context). As discussed
in the foregoing sections (see the discussion of figure \ref{fig:Triangle3new}
in section \ref{sec:Problem-description} and algo1 3.i and ,b; algo2
4.e and algo3b 2.c.v in section \ref{subsec:Formalized-algorithm-for}),
providing validation argumentation on each layer of abstraction (in
each step of iterative refinement) is a basic necessity. The high
level purpose-, operational design domain- and constraint analysis
forms the first level of concretization (open model iteration), based
on the \emph{high level goal}.

The high level operational design domain analysis and the high level
system hazards then form the basis for the system level hazard and
risk analysis. The system h\&r provides hazardous event characterization
and relation to the constraints and thereby identifies design drivers
without reducing design freedom (see 7. in the listing in section
\ref{subsec:Full-partition-of}).

Following this, the necessarily creative act of system design can
be approached in iterative, mutual dependent concretization of all
aspects ($T(R,C,P)$). This is done by applying \textbf{\emph{valid
deductive steps}}, addressing the deductive gap, as discussed in \ref{subsec:Deductive-gap,-validation}
and providing input for robustness and \emph{anti-fragility} (point
3 to 6 in the listing at end of section \ref{subsec:Full-partition-of})
by following the algorithms stated in section \ref{subsec:Formalized-algorithm-for},
triggering necessary refinement of the \emph{open models} and supporting
\emph{ongoing validation}.

With respect to preexisting components (e.g. radar or lidar sensors),
we suggest what we refer to as \textbf{\emph{contribution analysis}}:
the possibly to be used component from a lower level of abstraction
might be analyzed regarding the underlying assumptions, possible insufficiencies
and consequences for validation etc. The open models can then be refined
with respect to the related aspects (e.g. reflectivity coefficients,
presence of metallic structures in the operational design domain,
etc.) and design decisions can be taken on a higher level of abstraction,
based on this. The \emph{contribution analysis} steps are an important
input for design for validation and allow steering high level design
from a perspective of 'what can be validated and what can be built'.
This is necessary, as it makes no sense to design a system purely
top down, finally ending up with a solution that can either not be
build or hardly validated. 

\subsection{Contributions to the holistic evidence generation\label{sec:Contribution-of-E}}

We comment only briefly on the contributions to the holistic \textbf{\emph{evidence
generation cycle}} necessary for the validation of complex systems
operating in an open context (related to algo1 3.i and algo1 3.j,
algo2 4.e and algo3b 2.c.v in section \ref{subsec:Formalized-algorithm-for}).
The main contributions are from 
\begin{itemize}
\item system understanding
\item simulation
\item real world observation
\item continuous observation (pre- and post-release e.g. in the style of
a control and observation center)
\item continuous feedback to system understanding
\end{itemize}
It is important to note that all individual contributions can never
be used as silver bullet (i.e. singular approach for the complete
\emph{validation challenge}). Due to the characteristics of the problem
elaborated in the foregoing sections, every possible contribution
per se can never be complete. For complex systems operating in an
open context, the individual contributions necessarily need to be
mutually reinforced by combination with the others. Only closing the
whole cycle in a well balanced form across all aspects listed above
will be appropriate to address the \emph{validation challenge}. However,
the result will only be as good as the individual contributions. In
other words, excessively investing in singular aspects (e.g. simulation)
and neglecting others will be inappropriate.

As an example, we comment on obvious, however problematic statistics
based \emph{silver bullet} approaches in the sense of 'driving X hours
or miles' to demonstrate the validity of the system, which have initially
been discussed for autonomous driving, but never seriously been applied
(as \emph{silver bullet}). As should have become clear so far, the
characteristics of the \emph{validation challenge, }especially the
\emph{representativeness challenge} and the problems related to extremely
rare, but systematic manifestation of unacceptable losses, render
such an approach fundamentally intractable. Nevertheless, magic numbers
X could be proposed and even reduced in size by superficial statistic
arguments, e.g. about data fusion. Besides the fact that, due to the
aforementioned reasons already the starting point of this argumentation,
namely the magic number X, would be basically questionable, a conscientious
analysis shows that the basic problem of X being large would even
be irreducible for realistic systems within a \emph{silver bullet}
approach. For example, a perfect independence of fused data can not
ad-hoc be argued. However, demonstrating even a certain level of independence
already poses a problem even larger than X (see \cite{Butler1993},
especially section 5 for a detailed discussion). Statistic arguments
however, when being part of the holistic evidence cycle, have an important
contribution to validation. 

The general process sketched in section \ref{subsec:Overview-proc}
and formalized in section \ref{subsec:Formalized-algorithm-for},
forms a framework to handle the fragmentary, iteratively refined knowledge
about the operational design domain, the system design (including
design alternatives), and implementation (i.e. it addresses C, P and
R as well as the necessary validation argumentation across all levels).
This provides an adequate understanding of the open context total
system (e.g. ego vehicle and relevant part of the surrounding), the
interactions and mutual dependencies, as well as potential insufficiencies
(i.e. assumptions being temporarily or permanently invalid). From
this, necessary evidence contributions and hence validation and verification
activities can be derived. System understanding therefore is one basic
pillar of the \emph{holistic evidence generation cycle}. It addresses
subareas 1-4,7-11 and the fuzzy edge of ($\tilde{T}$+loss) in figure
\ref{fig:vennFull} and prepares metrics for related evidence generation. 

Another contribution comes from simulation. A full simulation of open
context systems is intractable. However, it has a strong contribution
in developing understanding on certain levels e.g. sensor insufficiencies
or behavior simulation. In addition, simulation is well suited for
scenario analysis (what if) and therefore can provide evidence for
the support of the worked out robust- and anti-fragile design, metrics,
recovery and degradation strategies etc. With other words, it strengthens
the arguments formed in the system understanding pillar.

Real world observation, on test tracks or real context, with a small
number of ego entities or connectivity based approaches across larger
fleets (pre- and post-release), on the other hand, can provide contributions
about aspects which can not easily be formally analyzed in detail
(system understanding) or simulated. This inability to simulate might
be due to sheer complexity being intractable (e.g. full ray-tracing
for sensors) or aspects unwittingly missing in $\bar{T}$. Real world
observation therefore is the only means able to address area 5 and
6 in figure \ref{fig:vennFull}. On the other hand, the \emph{representativeness
challenge} makes it hard to representatively sample the open context
for all events relevant for the complex system.

Therefore, being prepared for the unexpected and a continuous observation
in the sense of \emph{ongoing validation} (based on system understanding)
and continuous feedback to system understanding complete the holistic
cycle. The feedback allows to continuously confirming or refining
the system understanding, possibly triggering rework for improvement
(restarting next cycle iteration), as well as to strengthen the evidence
for the validity of the system.

\subsection{Contribution of from ISO26262 and SOTIF \label{sec:ContributionsIsoSotif}}

With respect to ISO26262 and SOTIF, one needs to distinguish between
what is requested (i.e. which problem is posed) and which part of
the posed problem is provided support for.

ISO26262 and SOTIF both are focused only on a part of the \emph{unacceptable
losses}, namely harm of persons related to E/E systems. All the other
aspects of a \emph{valid} product discussed in section \ref{subsec:Conception-of-validity}
are out of scope. 

An adequate understanding of the necessary functionality (of the item)
and the interaction with the environment and other items is requested
by ISO26262 (see. e.g. ISO26262 section 3.5), but regarded as to come
from external source. In other words, an adequate system understanding
and item definition (related to $\bar{T}$) is required, but generation
is not supported. There is a good support for addressing systematic
development as well as hardware failures, with \textbf{\emph{failure
(I)}} defined as\emph{ termination of the ability of an element, to
perform a function as required.} This is related to the crossover
between area 1 and 2 in figure \ref{fig:vennFull}. Broadly speaking,
given an adequate item definition, everything which follows therefrom
and can be addressed by verification (well defined starting point
and iterative crossover to well defined conretizations) is supported.
For non-complex systems in a defined context, this might be appropriate.
However, for complex systems operating in an open context, as argued
in the foregoing sections, a strong support especially for validation
issues would be mandatory. 

ISO26262 request a \textbf{\emph{safety validation (I)}} - \emph{assurance,
based on examination and tests, that the safety goals are sufficient
and have been achieved.} In addition, \textbf{\emph{functional safety
(I)}} is understood as \emph{absence of unreasonable risk }(i.e. related
to harm of persons)\emph{ due to hazards caused by malfunctioning
behavior of E/E systems}, with \textbf{\emph{malfunctioning behavior
(I)}} defined as \emph{failure or unintended behavior of an item with
respect to its design intent.} Unintended behavior ($T^{\dagger}\leftrightarrow\bar{T}$)
with respect to design intent $(T^{\dagger}\leftrightarrow\tilde{T}$+loss)
is related to area 5 in figure \ref{fig:vennFull}. Therefore, ISO26262
in fact already poses the full problem, however, as already stated
above, does not provide support for addressing all related aspects.

SOTIF is an approach to overcome the limitations of ISO26262 (currently
for driver assistance systems). The PAS (in preparation) states:
\begin{quote}
\emph{For some systems, which rely on sensing the external or internal
environment, there can be potentially hazardous} (related to harm
of persons)\emph{ behavior caused by the intended functionality or
performance limitation of a system that is free from the faults addressed
in ISO26262.}
\end{quote}
Hence, \textbf{\emph{safety of the intended functionality (I)}} is
defined as \emph{absence of unreasonable risk }(related to harm of
persons)\emph{ due to hazards resulting from functional insufficiencies
of the intended functionality or from reasonably foreseeable misuse
by persons }

As already discussed in the problem statement (section \ref{sec:Problem-description}),
SOTIF's \emph{functional insufficiency} relates to what we more generally
refer to as\emph{ insufficiency}, fundamentally related to invalid
assumptions. The SOTIF process - in contrast to the ISO26262 explicitly
addresses the iterative refinement of the functional- and system specification
with respect to \emph{insufficiencies} (i.e. invalid, possibly implicit
\emph{assumptions}). See e.g. ISO/PAS 21448's figure 9, flowchart
of the activities, appended for reference as figure \ref{fig:sotif9}
in the appendix. The step from activity 'start' to 'functional and
system specification' relates to \emph{sys$^{2}$val}'s high level
goal definition followed by the \emph{valid deductive step} supported
'well controlled concretization' of the aspects $P,R$ and $C$ of
the \emph{validation triangle} (i.e. iterative deduction of $\bar{T}$,
including \emph{contribution analysis steps}). Section 5 of the PAS
provides a listing of elements which should be part of the resulting
specification. The\emph{ sys$^{2}$val} approach sketched above provides
support for the generation of these and further elements necessary
for complex systems operating in open contexts.

When over-viewing the process of the PAS activities as depicted in
the flowchart, it is important to note that the focus actually is
on \emph{insufficiencies} (i.e. invalid \emph{assumptions}) and analysis
is based on system understanding. Triggering events should be regarded
as subordinated (with respect to \emph{insufficiencies}). They are
representative events stimulating the related \emph{insufficiencies}
and hence possibly leading to harm of persons. With subordinated we
refer to the fact that the fundamental basis is formed by analysis
of\emph{ insufficiencies} and not by triggering events. A real world
observation focused, triggering event based \emph{silver bullet} like
approach would be inappropriate, at least for complex systems operating
in open contexts. In the sense of the \emph{holistic evidence generation
cycle} (section \ref{sec:Contribution-of-E}), real world observation
and collection of triggering events in order to enrich and refine
system understanding however is an important part of a larger overall
approach. 

With this in mind we can regard area 1 of the flow chart as generation
of a collection of representative triggering events (the actual aim
is to achieve a good \emph{assumption} coverage) and in this sense
being related to the activities targeted at enlarging area 1 and the
\emph{assumption} \emph{monitoring} circular ring, especially area
4 of our figure \ref{fig:vennFull}. However, we, once again, refer
to the \emph{representativeness challenge}. Assuming triggering event
representativeness would be possible, the PAS flowcharts area 2, especially
the reiteration of the SOTIF process via 'functional modification',
relates to activities intended to transform the system such that subarea
of our area 4 is converted to area 1 in figure \ref{fig:vennFull}.

And finally, PAS flowcharts area 3, especially the reiteration via
'functional modification', relates to activities intended to transform
the system such that subarea of our area 5 is converted to area 1
in figure \ref{fig:vennFull}. A strong support in doing so, however,
is currently not part of the PAS.

It needs to be noted that the ISO/PAS 21448 is strongly influenced
by the state of the art approach for assistance systems, whereby statistic
methods play an important role. In fact, the ISO/PAS 21448 currently
is aimed solely for the application of level 2 (driver assistance)
systems. Preparation for an extension towards level 3+ is currently
in discussion.

Based on our problem statement and comparing the PAS activities to
the necessities discussed in the foregoing sections, it becomes clear
that the current draft of the ISO/PAS 21448 leaves open a gap with
respect to complex systems operating in an open context. In addition
to the discussed aspects in this section, \emph{ongoing validation}
and 'being prepared for continuous improvement' (\emph{anti-fragility})
play an important role for more complex systems validation and should
not be neglected when extending ISO/PAS 21448.

\section{Conclusion}

A fundamental understanding of the \emph{validation challenge} related
to design and operation of complex systems in open contexts is mandatory,
in order to be able to provide strategies for 'viable' validation
and approval of such systems. We gave a detailed problem statement,
formalization thereof and formalized algorithms for iterative development
and validation, based on this. Due to the complexity of the validation
issue, this formalization is regarded mandatory in order to ensure
that none of the many subtle details are neglected in development
process design. In addition, we provide a high level overview of a
practical, holistic development process (sys$^{2}$val) and comment
on the contributions from ISO26262 and current state of ISO/PAS 21448
(SOTIF).

From a practical point of view, basic challenges arise from managing
the large number of fragmentary and iteratively growing knowledge
(see all the interrelated aspects being part of the representation
on every level of abstraction $\{Rep_{i}\}$, eq. (\ref{eq:rep})).
The sheer mass and complex interrelation of the knowledge fragments
(besides the complex elicitation and evidence generation processes)
poses demands on knowledge engineering approaches and -tools, which
is well beyond the state of the art. 

Another basic challenge is due to the necessity of integrating fragmentary
and manifold evidences to an overall approval argumentation on the
one hand, and the actual validation goal ($\tilde{T}+loss$) being
a moving target strongly related to societal factors. As of today,
a unified theory of evidence integration (e.g. integrating probabilistic
and systematic aspects) is not available. This however would allow
to formalize the state of - and guide the only just started societal
discussion, and hence provide formalized and widely accepted \emph{quality
criteria} $\{Qc\}$ (see section \ref{subsec:Formalized-algorithm-for}). 

Solving these challenges is necessary to successfully demonstrate
safe operation and strict avoidance of fatal incidents in all day
use, which is mandatory to ensure societal acceptance of highly autonomous
systems operating in open contexts, on the long run and preventing
a \textquoteleft winter of autonomous systems\textquoteright .

\section{References}



\section{Appendix}

\begin{figure}[H]
\begin{centering}
\includegraphics[viewport=50bp 415bp 595bp 842bp,clip,width=1\columnwidth]{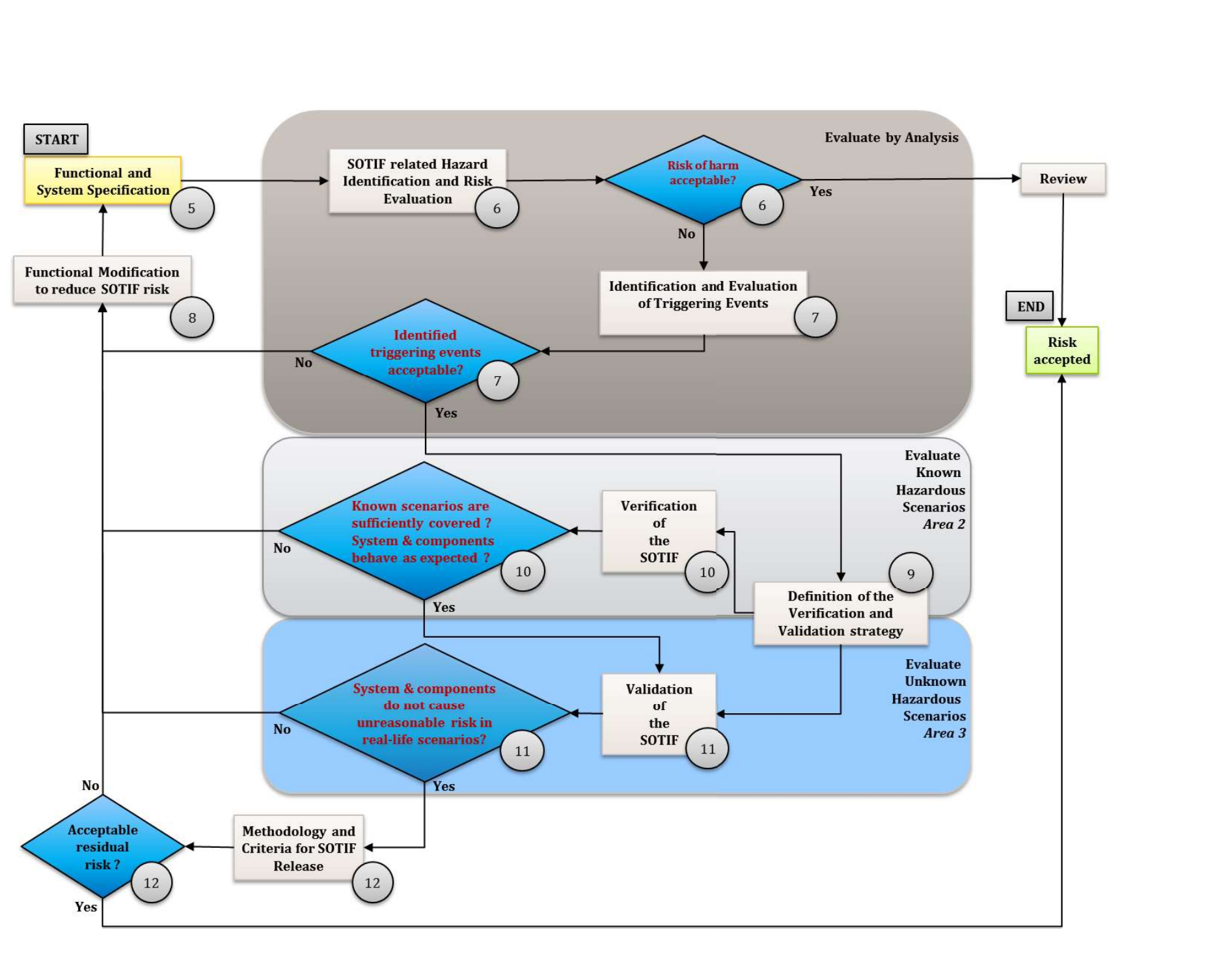}
\par\end{centering}
\caption{\label{fig:sotif9}Flowchart of the ISO/PAS 21448 activities (figure
9 in the draft version of ISO/PAS; changes to the final publication
might apply.)}
\end{figure}

\end{document}